\documentclass[12pt,notitlepage,aps,preprintnumbers,superscriptaddress,nofootinbib,aps,prd,floatfix]{revtex4-1}
\pdfoutput=1
\usepackage{graphicx}
\usepackage{dcolumn}
\usepackage{bm}
\usepackage{slashed}
\usepackage{physics}
\usepackage{extarrows}
\usepackage{multirow,array}
\usepackage{amsmath,amssymb,physics}
\usepackage{subfigure,rotating}
\usepackage{enumitem}
\setlist{nolistsep}
\usepackage{xcolor}
\usepackage{url}
\usepackage{slashed}
\usepackage{hyperref}
\definecolor{nicered}{rgb}{0.5,0.,0.}
\definecolor{nicegreen}{rgb}{0.,0.5,0.}
\definecolor{niceblue}{rgb}{0.,0.,0.5}
\hypersetup{colorlinks,citecolor=blue,linkcolor=blue,urlcolor=blue}
\usepackage{orcidlink}
\usepackage{braket}
\usepackage{colortbl}
\usepackage{gensymb}

\usepackage{comment}
\usepackage{booktabs}

\usepackage[normalem]{ulem}
\usepackage{cancel}

\begin{document}
\preprint{TTK-25-17,  P3H-25-045}

\title{Modelling top-quark decays in  $\boldsymbol{t\bar{t}t\bar{t}}$  production at the LHC}

\author{Manal Alsairafi \,\orcidlink{0009-0007-6966-9780}}
\email{manal.alsairafi@rwth-aachen.de}
\author{Nikolaos Dimitrakopoulos \,\orcidlink{0009-0005-9063-9039}}
\email{ndimitrak@physik.rwth-aachen.de}
\author{Malgorzata Worek\,\orcidlink{0000-0002-6495-9136}}
\email{worek@physik.rwth-aachen.de}
\affiliation{Institute for Theoretical Particle Physics and Cosmology, 
RWTH Aachen University, D-52056 Aachen, Germany}

\date{\today}

\begin{abstract}
We compare the fixed-order NLO QCD predictions for the $pp\to t\bar{t}t\bar{t}+X$ process in the $4\ell$ decay channel with the parton-shower based results obtained with the \textsc{Powheg} and MC@NLO matching methods. In the first case, NLO QCD corrections are consistently included in both the $t\bar{t}t\bar{t}$ production step and the decays of the four top quarks, preserving all spin correlations. In the second approach, higher-order effects in top-quark decays with approximate spin correlations are simulated in the \textsc{Pythia} parton-shower framework. Additionally, we analyse the impact of including the so-called matrix element corrections in top-quark decays in both parton-shower matched predictions. The comparison is performed at the integrated and differential fiducial cross-section level for the LHC  centre-of-mass energy of $\sqrt{s}=13.6$ TeV.
\end{abstract}

\maketitle


%
\section{Introduction}
%

The $pp\to t\bar{t}t\bar{t}$ process is a fascinating, yet relatively rare process that can be measured at the Large Hadron Collider (LHC). Precisely because it is a rare process, $t\bar{t}t\bar{t}$ production is sensitive to contributions from potential new physics. Many of the new physics scenarios predict effective $t\bar{t}t\bar{t}$ interactions as well as new processes such as $pp\to Y_{new} Y_{new}$ and $pp \to t\bar{t}Y_{new}$, where $Y_{new}$ denotes a new heavy resonance decaying into a $t\bar{t}$ pair, leading to the $t\bar{t}t\bar{t}$ final state. Any significant deviation from the Standard Model (SM) prediction for the $t\bar{t}t\bar{t}$ cross section could hint at these new particles or interactions. Consequently, good theoretical control of the $pp\to t\bar{t}t\bar{t}$ process is a fundamental requirement for the correct interpretation of possible new physics signals that may appear in this channel. In addition, top quarks decay into a bottom quark and a $W$ boson, and the $W^\pm$ bosons decay further into leptons or quarks. For the  $pp\to t\bar{t}t\bar{t}$ process, this can lead to complicated final states with multiple jets, including $b$-jets, multiple leptons, and missing transverse momentum due to neutrinos that are not detected at the LHC. These rich signatures allow for powerful cross-checks between different theoretical predictions to strengthen confidence in modelling of the SM process, but also bring significant experimental challenges from large backgrounds, especially from $pp \to t\bar{t}W^\pm$, $pp\to t\bar{t}W^{\pm}+jets$, $pp\to t\bar{t}Z/\gamma$,  $pp\to t\bar{t}H$ or other processes with multiple leptons and jets. 

At the LHC, both ATLAS and CMS measured the $pp\to t\bar{t}t\bar{t}$ cross section at a centre-of-mass energy of $\sqrt{s}=13$ TeV \cite{ATLAS:2023ajo,CMS:2023ftu}. Both measurements are reasonably consistent with the SM prediction, although they are still subject to significant statistical and systematic uncertainties. Over time, as more data is collected, the measurements will become more precise, and analyses will focus on advancing such cross-section measurements, improving background estimation techniques, and looking for possible outliers that might indicate new physics. In addition to the total cross section, differential measurements will be performed and compared to simulated theoretical predictions. These tests would help refine the modelling aspects for this process and highlight potential deviations that might point to new physics effects. 

With the high-luminosity upgrade of the LHC on the horizon, even more precise measurements of $t\bar{t}t\bar{t}$ production will become feasible. This must be accompanied by equally precise and accurate theoretical predictions that are under good control. Improved modelling may be key to matching the increased experimental precision. Modelling of the $pp\to t\bar{t}t\bar{t}+X$ process requires a combination of higher-order QCD calculations, careful handling of parton showers, and robust treatment of theoretical uncertainties. The complexity of the $t\bar{t}t\bar{t}$ final state makes accurate simulation particularly important, but also difficult, and should focus on distinguishing these rare events from various background processes.

Various theoretical predictions for the $pp\to t\bar{t}t\bar{t}$ process are currently available in the literature, focusing on the precision or accuracy of modelling different final states or both aspects simultaneously. Indeed, in the latter case, they not only take into account higher-order QCD effects, but also the proper treatment of spin correlations in top-quark decays, which is a necessary step to correctly describe many observables, including those sensitive to physics beyond the SM. Therefore, on the one hand, we have NLO QCD predictions for the $pp \to t\bar{t} t\bar{t}+X$ process with stable top quarks \cite{Bevilacqua:2012em,Alwall:2014hca,Maltoni:2015ena}. Theoretical predictions with the so-called complete NLO corrections are also available \cite{Frederix:2017wme}. In this case, various leading and subleading contributions, along with their higher-order effects, are consistently taken into account to form the complete NLO result. Recently, the calculation for this process has been carried out at the next-to-leading logarithmic level \cite{vanBeekveld:2022hty,vanBeekveld:2025ghw}. On the other hand, for realistic studies one has to consider a $12$-particle final state which consists of four $b$-jets and, depending on the decay channel, additional jets and/or charged leptons with missing transverse momentum. Also in this case, various theoretical predictions can already be found in the literature. These include the NLO parton shower-matched predictions for the $1\ell$ decay channel \cite{Jezo:2021smh}, as well as the fixed-order NLO QCD predictions in the narrow-width approximation for the $4\ell$ \cite{Dimitrakopoulos:2024qib} and $3\ell$ decay channels \cite{Dimitrakopoulos:2024yjm}. While these are important theoretical results, they exist as separate predictions for different decay channels. No attempt has yet been made to consistently compare these two approaches for the same decay mode for the $pp \to t\bar{t}t\bar{t}+X$ process. It would be beneficial to compare the fixed-order NLO QCD results with those in which higher-order corrections to top-quark decays are generated by parton showers. Such a comparison could assess to what extent parton showers can approximate higher-order effects in top-quark decays. It would also be able to verify the significance of NLO QCD corrections to angular observables, which are sensitive to top-quark spin correlations as well as to new physics effects. Furthermore, such a comparison could help identify the phase-space regions sensitive to parton-shower effects, describing collinear and soft-collinear emissions with leading logarithmic accuracy.

The purpose of this paper is to take a first step in this direction and compare the fixed-order NLO QCD predictions with parton-shower based results for the $pp\to t\bar{t}t\bar{t}+X$ process in the $4\ell$ channel, thus, concentrating on the following final state $\ell^- \Bar{\nu}_\ell \, \ell^+ \nu_\ell \, \ell^-
\Bar{\nu}_\ell \, \ell^+ \nu_\ell \, b \bar{b} b \bar{b}$, where $\ell^\pm=e^\pm,\, \mu^\pm$. For the fixed-order NLO QCD results, theoretical predictions from Ref.  \cite{Dimitrakopoulos:2024qib} will be used. For predictions based on parton showers, we will build on the results presented in Ref. \cite{Jezo:2021smh} for the $1\ell$ decay channel, that are based on the \textsc{Powheg} matching scheme, and generate new results for  the $4\ell$ decay channel from scratch. In addition, we will provide theoretical predictions for the $pp \to t\bar{t}t\bar{t}+X$ process matched to parton showers using the MC@NLO matching scheme. Finally, we will analyse the impact of including the so-called matrix element corrections in the top-quark decays in both predictions matched to parton showers.
    
The paper is organised as follows. In section \ref{description} we outline the computational setup used in our study as well as highlight the differences between various theoretical approaches. In section \ref{sec:results} we present our results. This section is divided into two parts: theoretical predictions for the integrated (fiducial) cross sections, and predictions for the various dimensionful and dimensionless observables that will be measured at the LHC. We summarise our results in section \ref{sec:outlook}, where we also provide our final conclusions.

%
\section{Description of the Calculation}
\label{description}
%
%

In this study we consider the production and decay of four top quarks at NLO in QCD for the LHC centre-of-mass energy of $\sqrt{s}=13.6$ TeV. We provide both fixed-order predictions and results obtained from matching the NLO QCD calculation for the $pp\to t\bar{t}t\bar{t}$ process to parton showers. We concentrate on the $4$-lepton decay channel (denoted as $4\ell$) 
\begin{equation}
    p p \rightarrow t\bar{t}t\bar{t} +X \rightarrow W^+W^- W^+ W^-  b \bar{b} b \bar{b} +X \rightarrow\ell^- \Bar{\nu}_\ell \, \ell^+ \nu_\ell \, \ell^-
    \Bar{\nu}_\ell \, \ell^+ \nu_\ell \, b \bar{b} b \bar{b} +X\,,
    \label{eq: 4lepton}
\end{equation}
where $\ell \in\{e,\mu\}$. The SM parameters are fixed according to the $G_\mu$ scheme, with the Fermi constant given by
\begin{equation}
     G_{ \mu}= 1.166 378 \cdot 10^{-5} \,{\rm GeV}^{-2} \,,
\end{equation}
and with masses for the electroweak gauge bosons set to 
\begin{equation}
    \begin{array}{ccc}
      m_W = 80.379  \, {\rm GeV} \,, &\quad \quad \quad
      & m_{Z}=91.1876  \,{\rm GeV} \,.
    \end{array}
  \end{equation}
The electromagnetic coupling, $\alpha$, is then evaluated according to
\begin{equation}
\alpha_{G_\mu}=\frac{\sqrt{2}}{\pi} 
\,G_F \,m_W^2  \left( 1-\frac{m_W^2}{m_Z^2} \right)\,.
\end{equation}
The fixed-order calculations are performed in the five-flavour scheme, which accommodates massless bottom quarks that are not 
decoupled from the Parton Density Functions (PDFs). The proton content is parametrised using the MSHT20 PDF sets \cite{Bailey:2020ooq} provided by the LHAPDF library \cite{Buckley:2014ana}. The top-quark mass is set to
\begin{equation}
    m_t = 172.5 \, {\rm GeV}.
  \end{equation}
We keep the Cabibbo-Kobayashi-Maskawa matrix diagonal throughout the calculation. The final-state jets are constructed with the anti-$k_T$ jet algorithm \cite{Cacciari:2008gp}, with a clustering parameter $R$ set to $R=0.4$. We require the presence of exactly four charged leptons and at least four $b$-jets. We impose the following cuts on the transverse momentum, the rapidity and the angular separation of the $b$-jets
\begin{equation}
\begin{array}{lll}
p_{T}(b)>25 \,{\rm GeV}\,,  
&\quad \quad\quad\quad\quad |y(b)|<2.5 \,, 
 &\quad \quad\quad \quad \quad
\Delta R_{bb}>0.4\,.
\end{array}
\end{equation}
In order to ensure well-observable isolated (charged) leptons in the central-rapidity region of the ATLAS and CMS detectors, we additionally require
\begin{equation}
\begin{array}{lll}
 p_{T}(\ell)>25 \,{\rm GeV}\,,    
 &\quad \quad \quad \quad\quad|y(\ell)|<2.5\,,&
\quad \quad \quad \quad \quad
\Delta R_{\ell
 \ell} > 0.4\,.
\end{array}
\end{equation}
Finally, we do not set any constraint on the missing transverse momentum coming from the four neutrinos. The applied cuts are motivated by the experimental analyses from ATLAS \cite{ATLAS:2023ajo} and CMS \cite{CMS:2023ftu}. 

As the main purpose of this article is to compare the fixed-order NLO QCD predictions with the NLO QCD results matched to parton showers, in what follows, we will focus on describing these methods and providing additional input parameters that have not yet been described, as well as pointing out the differences between various  modelling approaches.

%
\subsection{Fixed-order predictions}
%

The NLO QCD corrections to the $pp\to t\bar{t}t\bar{t}$ process in the $4\ell$ channel are calculated with the help of the narrow-width approximation (NWA) \cite{Dimitrakopoulos:2024qib}. In this approximation only Feynman diagrams with four top quarks are taken into account in the calculation. We neglect triple-, double-, single- and non-resonant top-quark contributions, and do not include the Breit-Wigner propagator for the top quarks or $W^\pm$ gauge bosons. The NLO QCD corrections are included for both the $t\bar{t}t\bar{t}$ production stage and in all top-quark decays. We present the results following the NWA-expanded prescription of the calculation (denoted as NWA$_{\rm exp}$), where the cross section is given by 
\begin{equation}
  d\sigma_{\rm exp}^{\rm NLO} = d\sigma_{\rm full}^{\rm NLO}\times
  \left( \frac{\Gamma_t^{1}}{\Gamma_t^{0}}\right)^4
  - d\sigma^{\rm LO}\times\frac{4(\Gamma^{1}_t
    -\Gamma_t^{0})}{\Gamma_t^{0}}\,.
\label{eq: NWA exp xsec}
\end{equation}
Here, $d\sigma_{\rm full}^{\rm NLO}$ is the expression for the full NLO QCD calculation, where $\Gamma_t$ is not treated in a perturbative manner
\begin{equation}
\label{nwa_notexp}
\begin{split}
d\sigma^{\textrm{NLO}}_{\rm full} & ~=
 d\sigma^{1} \times \frac{d\Gamma^{0}_{t}}
  {\Gamma_{t}^{1}}
  \times \frac{ d\Gamma^{0}_{\bar{t}}}{\Gamma_{t}^{1}}
 \times \frac{d\Gamma^{0}_{t}}
  {\Gamma_{t}^{1}}
  \times \frac{d\Gamma^{0}_{\bar{t}}}{\Gamma_{t}^{1}} \\[0.2cm]
 & ~+ d\sigma^{0} \times \frac{d\Gamma^{1}_{t}}
  {\Gamma_{t}^{1}}
  \times \frac{ d\Gamma^{0}_{\bar{t}}}{\Gamma_{t}^{1}}
 \times \frac{d\Gamma^{0}_{t}}
  {\Gamma_{t}^{1}}
  \times \frac{d\Gamma^{0}_{\bar{t}}}{\Gamma_{t}^{1}} \\[0.2cm]
  & ~+ d\sigma^{0} \times \frac{d\Gamma^{0}_{t}}
  {\Gamma_{t}^{1}}
  \times \frac{ d\Gamma^{1}_{\bar{t}}}{\Gamma_{t}^{1}}
 \times \frac{d\Gamma^{0}_{t}}
  {\Gamma_{t}^{1}}
  \times \frac{d\Gamma^{0}_{\bar{t}}}{\Gamma_{t}^{1}} \\[0.2cm]
  & ~+ d\sigma^{0} \times \frac{d\Gamma^{0}_{t}}
  {\Gamma_{t}^{1}}
  \times \frac{ d\Gamma^{0}_{\bar{t}}}{\Gamma_{t}^{1}}
 \times \frac{d\Gamma^{1}_{t}}
  {\Gamma_{t}^{1}}
  \times \frac{d\Gamma^{0}_{\bar{t}}}{\Gamma_{t}^{1}} \\[0.2cm]
  & ~+ d\sigma^{0} \times \frac{d\Gamma^{0}_{t}}
  {\Gamma_{t}^{1}}
  \times \frac{ d\Gamma^{0}_{\bar{t}}}{\Gamma_{t}^{1}}
 \times \frac{d\Gamma^{0}_{t}}
  {\Gamma_{t}^{\rm 1}}
  \times \frac{d\Gamma^{1}_{\bar{t}}}{\Gamma_{t}^{1}} \,,
  \end{split}
\end{equation}
where $\sigma^0,\, \sigma^1$ and $\Gamma^0,\, \Gamma^1$ denote LO and NLO contributions to the $t\bar{t}t\bar{t}$ production and top-quark decay, respectively. They are computed with the NLO MSHT20 PDF set and NLO input parameters. Treating the $W$ gauge bosons in the NWA we can further write 
\begin{equation}
\label{eq: NWA W}
d\Gamma_t = d\Gamma(t\to bW) \sum_{\ell\nu_\ell} \frac{d\Gamma(W\to \ell \nu_\ell)}{\Gamma_W}\,.
\end{equation}
We also generate the results with the LO top-quark decays only (denoted as NWA$_{\rm LO dec}$). This approach is motivated by the fact that the $pp \to t\bar{t}t\bar{t}$ process used for the parton-shower based results is also generated with the LO top-quark decays. The cross section for the NWA$_{\rm LO dec}$ case is defined as
\begin{equation} \label{NLOlodec}
      d\sigma_{\rm LO dec}^{\rm NLO} = d\sigma^{1}
      \times \frac{d\Gamma_t^{0}}{\Gamma_t^{0}}
      \times \frac{d\Gamma_{\Bar{t}}^{0}}{\Gamma_t^{0}}
      \times \frac{d\Gamma_t^{0}}{\Gamma_t^{0}}
      \times \frac{d\Gamma_{\Bar{t}}^{0}}{\Gamma_t^{0}}\,.
\end{equation}

The fixed-order results are produced within the \textsc{Helac-NLO} framework \cite{Bevilacqua:2011xh} that comprises \textsc{Helac-1Loop} \cite{Ossola:2008xq,vanHameren:2009dr,vanHameren:2010cp} and
\textsc{Helac-Dipoles} \cite{Czakon:2009ss,Bevilacqua:2009zn,Bevilacqua:2013iha}. They are stored in the modified \textsc{Les Houches Event Files} \cite{Alwall:2006yp} and converted to the more compact
\textsc{Root Ntuple Files} \cite{Antcheva:2009zz, Bern:2013zja}, which are then reweighed and analysed with the help of the \textsc{HEPlot} program \cite{Bevilacqua:HEPlot}.  

We employ the following decay widths for the unstable particles 
\begin{equation}
    \begin{array}{ccc}
     \Gamma_W = 2.0972 \, {\rm GeV}\,, 
     & \quad \quad  \Gamma_t^{\rm NLO} 
     = 1.3535983 \, {\rm GeV}\,,
     & \quad \quad  \Gamma_t^{\rm LO} 
     = 1.4806842 \, {\rm GeV}\,.
\end{array}
\label{eq: helac widths}
\end{equation}
We use a dynamical scale setting for the renormalisation scale, $\mu_R$, and  factorisation scale, $\mu_F$, given by  $\mu_R=\mu_F=\mu_0=E_T/4$, where $E_T$ is defined according to
\begin{equation}
\label{eq: ET definition}
    E_T = \sum_{i\in\left\{ t,\bar{t}, t, \bar{t}\right\}}
    \sqrt{m_i^2 + p_T^2(i)} \,.
\end{equation}
In Eq.~\eqref{eq: ET definition} the momenta of the top and anti-top quarks are reconstructed from their decay products. In order to estimate the theoretical uncertainties arising from neglected higher-order terms in the perturbative expansion, we use the $7$-point scale variation in which $\mu_R$ and $\mu_F$ are varied independently in the range:
\begin{equation}
\frac{1}{2} \mu_0 \le \mu_R\,, \mu_F \le 2 \mu_0\,, 
\quad \quad \quad  \quad \quad \quad
\frac{1}{2} \le \frac{\mu_R}{\mu_F} \le 2\,.
\end{equation}
This leads to the following seven pairs that are taken into account in the error estimation
\begin{equation}
\label{scan}
\left(\frac{\mu_R}{\mu_0}\,,\frac{\mu_F}{\mu_0}\right) = \Big\{
\left(2,1\right),\left(0.5,1  
\right),\left(1,2\right), (1,1), (1,0.5), (2,2),(0.5,0.5)
\Big\} \,.
\end{equation}
In practice, we look for the minimum and maximum of the obtained cross-section results to provide the final estimate of the theoretical error resulting from scale dependence.

%
\subsection{NLO calculation matched to parton showers}
%

The results are simulated using Monte Carlo (MC) event generators based on the $pp \to t\bar{t}t\bar{t}+X$ process, supplemented with parton-shower programs. The matrix element generation for the $pp\to t\bar{t}t\bar{t}+X$ process is performed either within the \textsc{Powheg-Box} framework \cite{Nason:2004rx,Frixione:2007vw,Alioli:2010xd} or with the help of the \textsc{MadGraph5${}_{-}$aMC@NLO} program employing the  \textsc{MC@NLO} matching procedure for matching NLO order QCD calculations with parton showers \cite{Frixione:2002ik}. NLO QCD theoretical predictions are generated with LO top-quark and $W$ gauge-boson decays. In both approaches the generated events are showered via \textsc{Pythia} 8.3 \cite{Bierlich:2022pfr} and analysed using the \textsc{Rivet} \cite{Buckley:2010ar, Bierlich:2019rhm} framework, that includes an interface to the \textsc{HepMC3} event record library \cite{Buckley:2019xhk} and to the \textsc{FastJet}  library for jet finding \cite{Cacciari:2005hq, Cacciari:2011ma}. In addition, the effects of taking into account the matrix element corrections in top-quark decays are investigated, see e.g. Refs.  \cite{Frixione:2023hwz,Frederix:2024psm}. Finally, hadronisation, multi-parton interactions and QED showers are not addressed in this work.

%
\subsubsection{The MC@NLO method}
%

For the \textsc{MC@NLO}-type matching the  $pp\rightarrow t \Bar{t} t \Bar{t}$ process at NLO in QCD is generated with the help of the \textsc{MadGraph5${}_{-}$aMC@NLO} Monte Carlo program  \cite{Alwall:2014hca}. The renormalisation, factorisation and matching scales are chosen at this point based on the on-shell top-quark momenta. The dynamical scale setting $\mu_R=\mu_F=\mu_0 = E_T/4$ is employed, where the value of $E_T$ is computed according to 
\begin{equation}
    E_T = \sum_{i\in \left\{t, \Bar{t}, t, \Bar{t}, j \right\}}\sqrt{m_i^2 + p_T^2(i)} \,.
\label{eq: ET NLO+PS}
\end{equation}
The $\mu_R$ and $\mu_F$ scales are varied according to Eq. \eqref{scan}. The matching in the MC@NLO scheme depends crucially on the choice of the initial shower scale $\mu_Q$, which controls the available phase space for subsequent parton shower emissions and therefore can have a strong impact on the shower evolution.  We keep the \textsc{MadGraph5${}_{-}$aMC@NLO}  default choice of $\mu_Q=E_T/2$. We study the dependence on this scale by varying it by a factor of 2 up and down around the central value $\mu_Q$. The LO decays of the top quarks and the $W$ bosons are included via \textsc{MadSpin} \cite{Artoisenet:2012st}. The program in its default mode, which we retain, follows the approach described in Ref. \cite{Frixione:2007zp}. The decay widths of the unstable particles are calculated in the \textsc{MadSpin} framework from the input masses and couplings at LO accuracy
\begin{equation}
\begin{array}{cc}
     \Gamma_W = 2.0448 \, {\rm GeV}\,, & 
     \quad \quad \quad \quad \quad  \Gamma_t 
     = 1.4807 \, {\rm GeV}\,.
\end{array}
    \label{eq: madspin widths}
\end{equation}
As a result, we get the following branching ratio for the $W$ gauge boson
\begin{equation}
     {\cal BR}(W\rightarrow \ell \nu) = \sum_{i=1}^{3} 
    \frac{\Gamma (W \to \ell_i \nu_i )}{\Gamma_W} = 0.3332844\,.
\end{equation}
%

%
\subsubsection{The POWHEG method}
%

The \textsc{Powheg} results are generated following the implementation presented in Ref. \cite{Jezo:2021smh}. Similar to the MC@NLO approach, the  $pp\to t \bar{t} t\bar{t}$ process is generated at the NLO QCD level, while the top-quark decays are considered only at LO. More precisely, we use the method described in Ref. \cite{Frixione:2007zp} to preserve the spin correlations and introduce a smearing of top-quark and $W$ boson virtualities according to the Breit-Wigner distributions. The same dynamical scale setting is used for $\mu_R$ and $\mu_F$ as given in Eq. \eqref{eq: ET NLO+PS}. The matching procedure in the \textsc{Powheg} approach relies on the following two damping parameters: $h_{\rm damp}$ and $h_{\rm bornzero}$, see e.g. Ref. \cite{FebresCordero:2021kcc}. The $h_{\rm damp}$ parameter can either be set to a constant value or to a function of the underlying Born kinematics. On the other hand, $h_{\rm bornzero}$ is a dimensionless parameter. Both damping parameters are used to define a jet-function that classifies the soft and hard real contributions. We follow Ref. \cite{Jezo:2021smh} and set these parameters to 
\begin{equation}
        h_{\rm damp} = \frac{E_T}{4}\,, 
        \quad \quad \quad \quad h_{\rm bornzero} = 5\,.
\end{equation}
The impact of different choices of damping parameters is estimated by the $5$-point matching-scale variation, which corresponds to the calculation of the following contributions
\begin{equation}
    (h_{\rm damp},h_{\rm bornzero}) = \left\{ \left(\, \frac{E_T}{4}, 5 \,\right),
    \left(\, \frac{E_T}{4}, 2 \,\right),
    \left(\, \frac{E_T}{4}, 10 \,\right),
    \left(\, \frac{E_T}{8}, 5 \,\right),
    \left(\, \frac{E_T}{2}, 5 \,\right)
    \right\}\,.
\end{equation}
In addition, the initial shower scale is assigned to the transverse momentum of the extra emission generated in the event. 

The top-quark and $W$ gauge boson decay widths are evaluated at NLO in QCD taking into account finite-width effects of the $W$ gauge boson. Their values are given by
\begin{equation}
    \begin{array}{cc}
     \Gamma_W = 2.0972  \, {\rm GeV}\,, & \quad \quad \quad \quad  \Gamma_t = 1.33254\,  {\rm GeV}\,.
\end{array}
    \label{eq: powheg widths}
\end{equation}
The branching ratio for the leptonic decay of the $W$ gauge boson is set to 
\begin{equation}
    {\cal BR}(W \rightarrow \ell \nu) = 0.325036\,.
\end{equation}
It is worth mentioning here that the branching ratio and the $W$ gauge-boson width are independent parameters in the \textsc{Powheg} framework. Changing the value of $\Gamma_W$ will have a small impact on the results, since it only enters the Breit-Wigner distributions, which are responsible for the momentum mapping between on-shell and off-shell momenta  of the $W$ decay products. The same applies to the dependence on the top-quark width $\Gamma_t$.

%
\subsubsection{Matrix Element Corrections} \label{mec}
%

To correct for lack of higher-order QCD effects in top-quark decays in the parton-shower based computations, we employ the Matrix Element Corrections (MEC). We apply the MEC according to the prescription given in Ref. \cite{Frixione:2023hwz}. This allows us to avoid the double-counting problem, i.e. generation of emissions already accounted for at the matrix-element level. Specifically, the following settings are employed in our studies 
\begin{verbatim}
    SpaceShower:MEcorrections = off
    TimeShower:MEcorrections = on
    TimeShower:MEextended = off
\end{verbatim}
where \texttt{SpaceShowers} and \texttt{TimeShower} refer to initial- and final-state radiation, respectively. The \texttt{TimeShower:MEextended} flag, which is responsible for employing the  MEC also for $1\ \to n$ and $2 \to  n$ processes, is switched off. With this flag on, the  \textsc{Pythia} program would attempt to approximate the MEC for the whole  $pp \to t\bar{t}t\bar{t}$ processes for which no exact MEC exist. Enabling the extended MEC would therefore mean that the approximate MEC would be used also in the production phase. This is a situation we would like to avoid as it would lead to double counting, since NLO corrections to the production process are already incorporated in the calculation. When this flag is off, only the original splitting kernel for the emission in the $t\rightarrow b W$ decay is corrected by means of the $t\rightarrow b Wg$ tree-level matrix element. It should be noted, however, that the double-counting problem can only occur in MC@NLO type simulations. In the case of \textsc{Powheg}, we could also include the MEC in the production stage of the $pp\to t\bar{t}t\bar{t}$ process, since the hardest emission in the production is always guaranteed to be generated prior to showering. In this case, the impact of the production MEC would therefore be limited to only small $p_T$ regions, where the cross section is Sudakov suppressed and the emissions generated by \textsc{Powheg} do not occur. As a result, the production MEC are essentially negligible in \textsc{Powheg} and may or may not be included. Nevertheless, we decided to use the same settings for the \textsc{Powheg} and MC@NLO matching results.  

In addition, one can switch on the MEC to emissions after the first one via the following flag
\begin{verbatim}
    TimeShower::MEafterFirst = on.
\end{verbatim}
We leave this flag enabled as it is the default setting in \textsc{Pythia}. This would correspond, for example, to two consecutive corrected emissions from the $b$ quark, i.e. to such a process $t\to Wbg \,(g \to gg) \to Wbgg$.  We refer to this setup as $\oplus{\rm MEC}$.  However, we also investigate the case where the use of the MEC is not allowed after the first emission. This can be achieved by setting
\begin{verbatim}
    TimeShower::MEafterFirst = off.
\end{verbatim}
We refer to this setting as \textbf{$\oplus{\rm MEC^{1st}}$}. We expect the difference between $\oplus{\rm MEC}$ and \textbf{$\oplus{\rm MEC^{1st}}$} to be very small. We plan to check this for one type of matching, namely the MC@NLO matching.

%
\subsection{Differences in the various approaches}
%

Below we highlight the differences in the approaches described so far. Leaving aside the fact that the NLO plus parton shower results provide predictions with (soft/collinear) emissions included to all orders via repeated branching of quarks and gluons, here we will rather focus on the differences that arise in the calculations of the $pp \to t\bar{t}t\bar{t}+X$ process in the $4\ell$ decay channel. Such differences contribute to the systematic uncertainties for this process.

In the case of fixed-order NLO QCD predictions, following the $\rm NWA_{exp}$ approach, higher-order QCD corrections are consistently included in both the production and decay stages of the $pp \to t\bar{t}t\bar{t} + X$ process. Furthermore, spin correlations at the NLO level are preserved throughout the calculation. On the other hand, in the ${\rm NWA}_{\rm LOdec}$ results only LO top-quark decays are present and therefore only LO spin correlations are considered. Finally, although in the parton-shower based predictions the higher-order QCD corrections in top-quark decays are approximately taken into account, formally spin correlations are only LO accurate. Since the kinematics of the top-quark decay products are influenced by spin correlations, it is important to investigate the effects of higher-order corrections in the decays and to examine precisely to what extent parton showers can mimic these effects in order to fully understand the obtained results.

Another difference concerns the choice of the scale setting  used for $\mu_R$ and $\mu_F$. In the fixed-order approach, we define the dynamical scale setting via Eq.~\eqref{eq: ET definition}, where the additional light jet from the real emission part of the NLO calculation, if present, is excluded from the sum.  In contrast, the scale setting defined in Eq.~\eqref{eq: ET NLO+PS}, used in both parton-shower based computations, includes the additional light jet. We note here that  it is not possible to achieve  one-to-one correspondence in the scale setting between the fixed-order and parton-shower based predictions. In the former case, the scale setting is based on the observed final states (leptons and jets), where the momenta of the top and anti-top quarks are reconstructed from their decay products in the fiducial phase-space regions. In the latter case, the scale setting is based solely on the stable top quarks and extra jet. Unlike the fixed-order case, this additional jet can only come from the production stage of the process. Thus, different scaling settings can be considered as part of the systematic uncertainties.

A further distinction comes from the fact that the fixed-order NLO calculations depend on the chosen values of the $t$-quark and $W$-boson widths, as seen in Eq.~\eqref{eq: NWA exp xsec} and  
Eq.~\eqref{eq: NWA W}. The cross section for the parton-shower matched calculations, is instead given in terms of the production cross section times the corresponding branching ratios
\begin{equation}
    d\sigma^{\rm NLO+PS}_{4\ell} = d\sigma^{\rm NLO}_{pp\rightarrow t \Bar{t} t \Bar{t}}\times [{\cal BR}(t\rightarrow W^+ b)\, {\cal BR}(\Bar{t}\rightarrow W^- \Bar{b})]^2 \times 16\, [{\cal BR}(W^\pm \rightarrow \ell\nu)]^4\,, \\[0.2cm]
    \label{eq: 4lepton NLO+PS xsec}
\end{equation}
where the factor 16 accounts for all the combinations of the leptonic $W^\pm$ decays with $\ell=e,\, \mu$ and ${\cal BR}(W^\pm \rightarrow \ell\nu)$ refers to a given lepton flavour. Since the CKM matrix is kept diagonal in our calculations, we can write ${\cal BR}(t\rightarrow W^+ b)={\cal BR}(\Bar{t}\rightarrow W^- \Bar{b})=1$. Consequently, the dependence of $d\sigma^{\rm NLO+PS}_{4\ell}$ on the top-quark width vanishes. On the contrary, we are sensitive to ${\cal BR}(W^\pm \rightarrow \ell\nu)$. In \textsc{MadSpin}, all branching ratios are calculated with LO accuracy based on the given input parameters. On the other hand, in \textsc{Powheg} ${\cal BR}(W^\pm \rightarrow \ell\nu)$ is an independent input parameter which we set to the NLO value. Although the difference between the total leptonic $W$ boson branching ratio at LO and NLO is only $2.5\%$, the cross section depends on ${\cal BR}(W^\pm \rightarrow \ell\nu)$ quadruply. This increases the cross section by $\sim 10\%$ when LO branching factors are used instead. To consistently compare both approaches, we globally rescaled the results obtained within the \textsc{Powheg} framework by 
\begin{equation}
    \left( \frac{{\cal BR}^{\rm LO}(W \rightarrow \ell \nu)}{{\cal BR}^{\rm NLO}(W \rightarrow \ell \nu)}\right)^4 =\left( \frac{1/3}{{\cal BR}^{\rm NLO}(W \rightarrow \ell \nu)}\right)^4 = 1.106087\,. 
\end{equation}

The final conceptual difference between the fixed-order computations and the parton-shower matched predictions concerns the definition of the $b$-jets in the final state. The fixed-order calculations are performed in the 5-flavour scheme with massless bottom quarks. We employ the charge-blind $b$-jet tagging, see e.g. Ref. \cite{Bevilacqua:2021cit}. In practice the following set of recombination rules is used 
\begin{equation}
bg \to b\,, \quad \quad \bar{b} g \to \bar{b}\,, \quad \quad 
b\bar{b} \to g\,, \quad \quad bb \to g\,, \quad \quad
\bar{b}\bar{b}\to g\,,
\end{equation}
which makes the jet-flavour definition infrared safe at NLO in QCD. On the contrary, in the parton-shower matched simulations the $b$-quark is treated as a massive quark and the collinear divergence in the $g\to b\bar{b}$ splitting is regulated by the mass of the bottom quark. Thus, in this case a $b$-jet is always defined as a jet that has at least one $b$-quark among its constituents. Note here that this is actually closer to what is done experimentally, where the definition of $b$-jets is based on the presence of intermediate heavy long-lived $b$-flavoured hadrons decaying at displaced vertices. 

We would also like to add here that precise identification of $b$-jets can pose serious infrared and collinear safety problems for predictions beyond NLO in QCD. In such a case, the so-called flavoured anti-$k_T$ jet algorithms should be used instead, see e.g. Refs. \cite{Czakon:2022wam,Gauld:2022lem,Caola:2023wpj} and the comparative study presented in Ref. \cite{Behring:2025ilo}. Moreover, the flavoured anti-$k_T$ jet algorithms might be applied to parton-shower based predictions to check whether the obtained cross-section results contain large logarithms of the type $\log(m_b/p_{T,\,b})$. To ensure the robustness of our parton-shower based predictions, we also generated the integrated and differential cross-section predictions using a flavoured IR safe anti-$k_T$ jet algorithm that is available as a \textsc{FastJet} Plugin \footnote{\texttt{https://github.com/jetflav/IFNPlugin}}. All the differences at the integrated and differential cross-section level that we found were significantly lower than the corresponding theoretical uncertainties obtained using a scale variation procedure.

%
\section{Numerical results}
\label{sec:results}
%

In our study of the $pp\to t\bar{t}t\bar{t}+X$ process in the $4\ell$ decay channel we require the presence of exactly four charged leptons and at least four $b$-jets in the final state. In practice, however, for the fixed-order NLO QCD calculation we can only have up to five $b$-jets. On the contrary, in the parton-shower matched  predictions we can easily observe up to  ten or even eleven $b$-jets.  Moreover, in this case the four $b$-jets with the highest $p_T$ do not always correspond to the $b$-jets  generated in the top-quark decays. Understanding the origin of $b$-jets is particularly important when the identification of $b$-jets from top-quark decays is required, which can be the case, for example, when extrapolating the fiducial measurement to the full phase space in order to unfold to stable top quarks, or when the $pp\to t\bar{t}t\bar{t}+X$ process is treated as one of the background processes in the search for new physics. Therefore, in addition to the general comparison of the four hardest $b$-jets from the fixed-order predictions with those from the NLO plus parton-shower simulations, in the latter case we will also attempt to identify the $b$-jets originating directly from the top-quark decays. This can be achieved by exploiting the so-called mother-daughter history available within the \textsc{Pythia} framework from the production vertex to a given final state, see e.g. Ref. \cite{Bierlich:2022pfr}. The $b$-jets produced at the matrix-element level can originate not only from the resonant top quarks, but also from the real-emission part of the NLO QCD calculations, e.g. from the following subprocesses $gb/bg \to t\bar{t}t\bar{t} \,b$ and $g\bar{b}/\bar{b}g \to t\bar{t}t\bar{t}\, \bar{b}$. However, such contributions are small, of the order of $0.2\%$, and are therefore disregarded. Thus, we identify  a $b$-jet from the matrix-element calculation as a jet that contains a bottom quark coming either from $t$ or $\bar{t}$. Any $b$ quark originating from a $g\rightarrow b \Bar{b}$ splitting is excluded. We then require that, among all $b$-jets in the event, there are always exactly four such $b$-jets present, and only these events are taken into account when constructing integrated and differential cross-section predictions labelled  as the NLO plus parton shower predictions (NLO+PS) with the $b$-jet identification. 

%
\subsection{Integrated cross-section predictions}
%

\begingroup
\setlength{\tabcolsep}{8pt} 
\renewcommand{\arraystretch}{1.2} 
\begin{table}[t!]
\centering
\begin{tabular}{ccccccc}
  \hline\hline
\textsc{Theoretical Approach} & $\sigma$ [ab] & $+\delta^{\rm scale}$&$-\delta^{\rm scale}$& +$\delta^{\rm match}$ & -$\delta^{\rm match}$ & $\sigma/\sigma_{\rm NWA_{exp}}$ \\
   \hline\hline
\multicolumn{7}{c}{\textsc{Fixed-Order NLO}} \\
   \hline\hline
${\rm NWA_{exp}}$ & 5.170(3) & +12\% & -20\% & - & - & 1.0000 \\
${\rm NWA_{LO_{dec}}}$ & 5.646(3)  & +22\% & -23\% & - & - & 1.0921 \\
   \hline\hline
\multicolumn{7}{c}{\textsc{NLO+PS without $b$-Jet Identification}} \\
   \hline\hline
${\rm MC@NLO+PS}$ & 4.891(8) & +15\% & -21\% & +1.1\% & -0.7\% & 0.9460 \\
${\rm MC@NLO+PS\oplus MEC^{1st}}$ & 5.081(8) & +15\% & -21\% & +0.9\% & -0.6\% & 0.9828 \\
${\rm MC@NLO+PS\oplus MEC}$ & 5.098(8) & +15\% & -21\% & +0.7\% & -0.5\% & 0.9861 \\
\hline
${\rm POWHEG+PS}$  & 4.820(6) & +15\% & -21\% & +0.0\% & -0.5\% & 0.9323\\
${\rm POWHEG+PS\oplus MEC}$& 5.024(6) & +15\% & -21\% & +0.0\% & -0.5\% & 0.9718  \\
   \hline\hline
\multicolumn{7}{c}{\textsc{NLO+PS with $b$-Jet Identification}} \\
   \hline\hline
${\rm MC@NLO+PS}$ & 4.708(7) & +15\% & -21\% & +0.8\% & -0.4\% & 0.9106 \\
${\rm MC@NLO+PS\oplus MEC^{1st}}$ & 4.910(8) & +15\% & -21\% & +0.6\% & -0.4\% & 0.9497 \\
${\rm MC@NLO+PS\oplus MEC}$ & 4.923(8) & +15\% & -21\% & +0.5\% & -0.5\% & 0.9522 \\
\hline
${\rm POWHEG+PS}$  & 4.656(6)   & +15\% & -21\% & +0.1\% & -0.2\% & 0.9006  \\
${\rm POWHEG+PS\oplus MEC}$& 4.865(6) & +15\% & -21\% & +0.0\% & -0.2\% & 0.9410  \\
   \hline\hline
\end{tabular}
    \caption{\textit{Integrated fiducial cross-section predictions
    at NLO in QCD for the $pp\rightarrow t \Bar{t}t \Bar{t}+X$
    process in the $4\ell$ channel. We present the fixed-order
    results and predictions matched to parton shower. For the
    fixed-order case we provide the results for $\sigma^{\rm
    NLO}_{\rm NWA_{exp}}$ and for $\sigma^{\rm NLO}_{\rm
    NWA_{LO_{dec}}}$. The parton-shower based results are
    presented without and with the $b$-jet identification as well as for the various MEC settings. The scale uncertainties, $\delta^{\rm scale}$, and parton-shower matching uncertainties, $\delta^{\rm match}$, are also reported.}}
    \label{table: 4lepton integrated}
\end{table}
\endgroup

We start by presenting and comparing the integrated fiducial cross-section predictions for the $pp \to t\bar{t} t\bar{t}+X$ process in the $4\ell$ decay channel together with their corresponding theoretical uncertainties. In particular, we provide theoretical uncertainties estimated from independent variations of the renormalisation and factorisation scales. For the parton-shower based predictions we also show the parton-shower matching uncertainties. For the results obtained with the \textsc{MC@NLO} matching the latter uncertainties are estimated by a variation of the initial shower scale $\mu_Q$, whereas for the \textsc{Powheg} case the variation of the two damping parameters is performed instead. 

Our findings  are summarised in Table \ref{table: 4lepton integrated}. Several statements are in order now. Firstly, all parton-shower based predictions are smaller than the fixed-order results. Indeed, in the presence of multiple shower emissions the parton shower allows the $b$-jets to evade kinematical bounds that would be otherwise present at the fixed-order case. Moreover, omitting NLO QCD corrections in top-quark decays increases not only the value of the integrated cross section by about $9\%$, but also the size of the theoretical uncertainties arising from scale dependence from $20\%$ to $23\%$. The NLO plus parton shower predictions without any MEC and $b$-jet identification differ from the default  $\sigma^{\rm NLO}_{{\rm NWA_{exp}}}$ fixed-order result by $-5\%$ and $-7\%$ for the \textsc{MC@NLO} and \textsc{Powheg} prediction, respectively. Nevertheless,  these differences are well within the quoted theoretical uncertainties, which are of the order of $21\%$ for the parton-shower based predictions. 

Secondly, when the $b$-jet identification is used, about $3\%-5\%$ more events are vetoed. This simply means that there might be $b$-jets from parton showers that undergo selection cuts and easily mimic the $b$-jets from the top-quark decays. The latter, due to multiple emissions of light jets, no longer pass the required cuts. This leads to differences of up to $10\%$ with respect to $\sigma^{\rm NLO}_{\rm NWA_{exp}}$.

Thirdly, we note that including the MEC in the top-quark decays increases the integrated fiducial cross section by about $4\%-5\%$ regardless of whether the $b$-jet identification is used. Moreover, as expected the difference between  the two results \textbf{$\oplus{\rm MEC^{1st}}$}  and \textbf{$\oplus{\rm MEC}$} is less than $1\%$. Overall, the  inclusion of the MEC leads to better agreement with the fixed-order $\rm NWA_{exp}$ result, highlighting the importance of the MEC in parton-shower based predictions. 

We can further observe that the NLO QCD parton-shower based predictions for the $pp \to t\bar{t}t\bar{t}+X$ process in the $4\ell$ channel, as obtained with the \textsc{MC@NLO} and \textsc{Powheg} matching methods, are in a rather good agreement. 
\begin{figure}[t]
    \centering
    \includegraphics[width=0.9\linewidth]{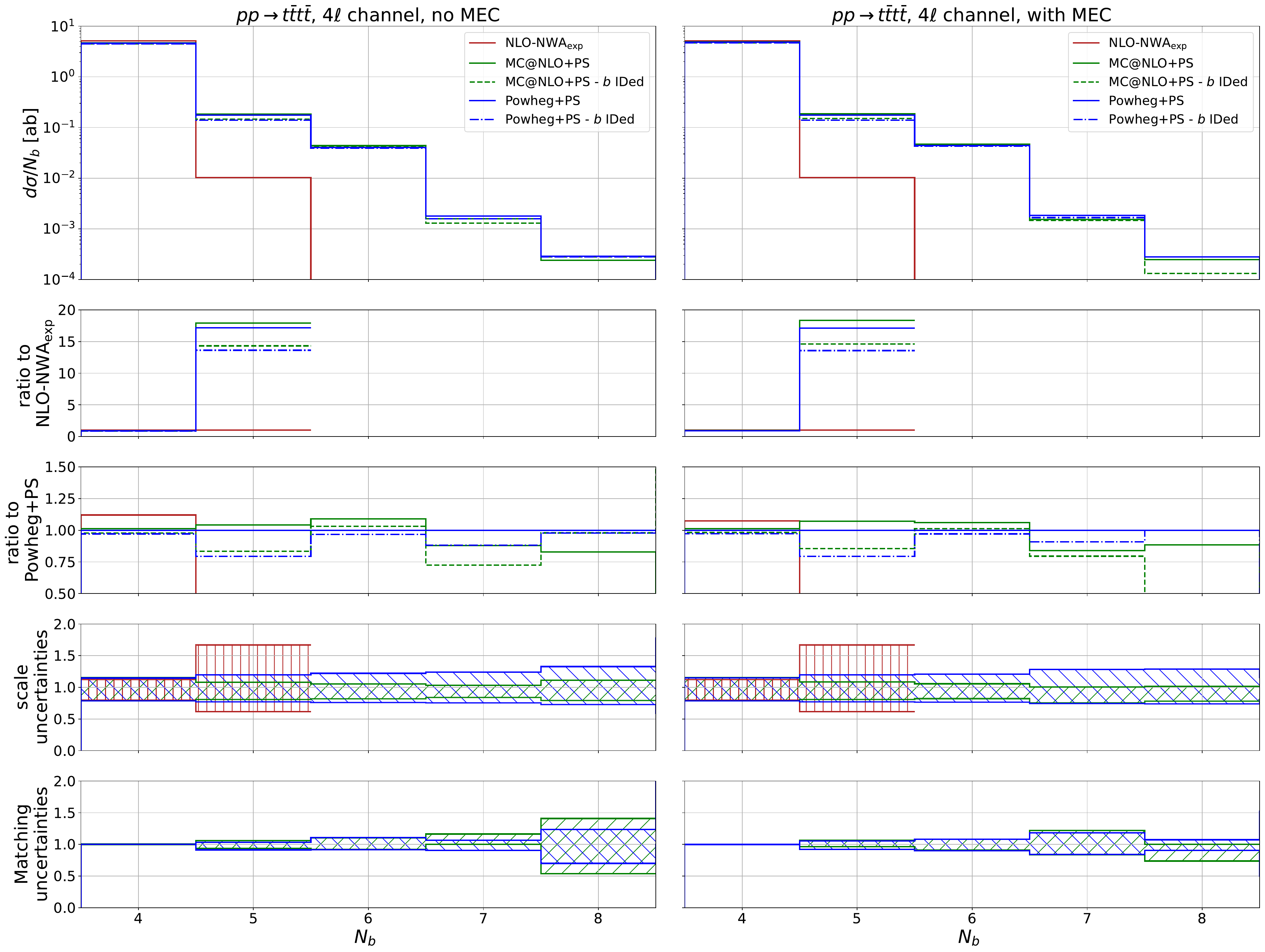}
    \caption{\textit{Integrated fiducial cross-section predictions
    at NLO in QCD for the $pp\rightarrow t \Bar{t}t \Bar{t}+X$
    process in the $4\ell$ channel as a function of the number of $b$-jets present. The parton-shower based results are given without (left) and with the MEC (right). Predictions without (solid line) and with (dashed line) the $b$-jet identification are also shown. The second panel from the top provides the ratio to the fixed-order results. The third panel shows the ratio to the \textsc{Powheg} predictions. The fourth panel presents the scale uncertainties, and the bottom panel depicts the parton-shower matching uncertainties.}}
    \label{fig:4lepton-n_b}
\end{figure}

It is interesting to see the results for the particular number of the $b$-jets. To this end, in Figure \ref{fig:4lepton-n_b} we present the integrated fiducial cross-section predictions at NLO in QCD as a function of the number of final state $b$-jets. Each bin indicates the cross-section result with the exact number of $b$-jets. The parton-shower based results are given without and with the MEC  as well as without and with the $b$-jet identification. The second panel from the top shows the ratio to the fixed-order predictions.  The third panel presents the ratio to the \textsc{Powheg} results. This way we can better observe the spread of the parton-shower based results  for high $b$-jet multiplicities. The fourth panel depicts the theoretical uncertainties obtained by scale variation, whereas the lower panel provides the parton-shower matching uncertainties. Since the fixed-order NLO QCD predictions only include up to $5$ $b$-jets, the first bin is NLO accurate, the second bin LO accurate, and starting from $N_b=6$ the $b$-jets are described solely by the shower evolution. The cross section with exactly $4$ $b$-jets gives the dominant contribution to the integrated cross section for the $pp\to t\bar{t}t\bar{t}+X$ process in the $4\ell$ channel. In this bin, the NLO plus parton-shower predictions are smaller than the NLO QCD fixed-order results by about $10\%$. This difference is, however, well within the shown  scale uncertainties, which are of the order of $20\%$. On the other hand, for the case with exactly $5$ $b$-jets the parton-shower based predictions are up to $19$ times larger than the fixed-order result. Moreover, the increased theoretical uncertainties in the latter case reflect its LO nature. In addition, we can observe that the various NLO plus parton showers results  agree with each other within the given scale uncertainties. As the number of the $b$-jets increases, the size of the scale and matching uncertainties increases as well. For a large number of $b$ jets, they become of similar size, especially if the MEC are not applied.

Before we turn to the differential cross-section results, let us  briefly note that in this study we do not analyse the PDF uncertainty. This is because they have already been discussed in detail in Ref. \cite{Dimitrakopoulos:2024qib} for the fixed-order NLO QCD predictions. Nevertheless, for the sake of completeness, we would  like to state that the PDF uncertainties for the $pp \to t\bar{t}t\bar{t}+X$ process in the $4\ell$ channel are up to $2\%$ for the NNPDF3.1 PDF set \cite{NNPDF:2017mvq}, of the order of $3\%-4\%$ for MSHT20 (our default PDF set) and up to $5\%-6\%$ for the CT18 PDF set \cite{Hou:2019efy}.

%
\subsection{Differential cross-section predictions}
%

\begin{figure}[t]
    \centering
    \includegraphics[width=0.9\linewidth]{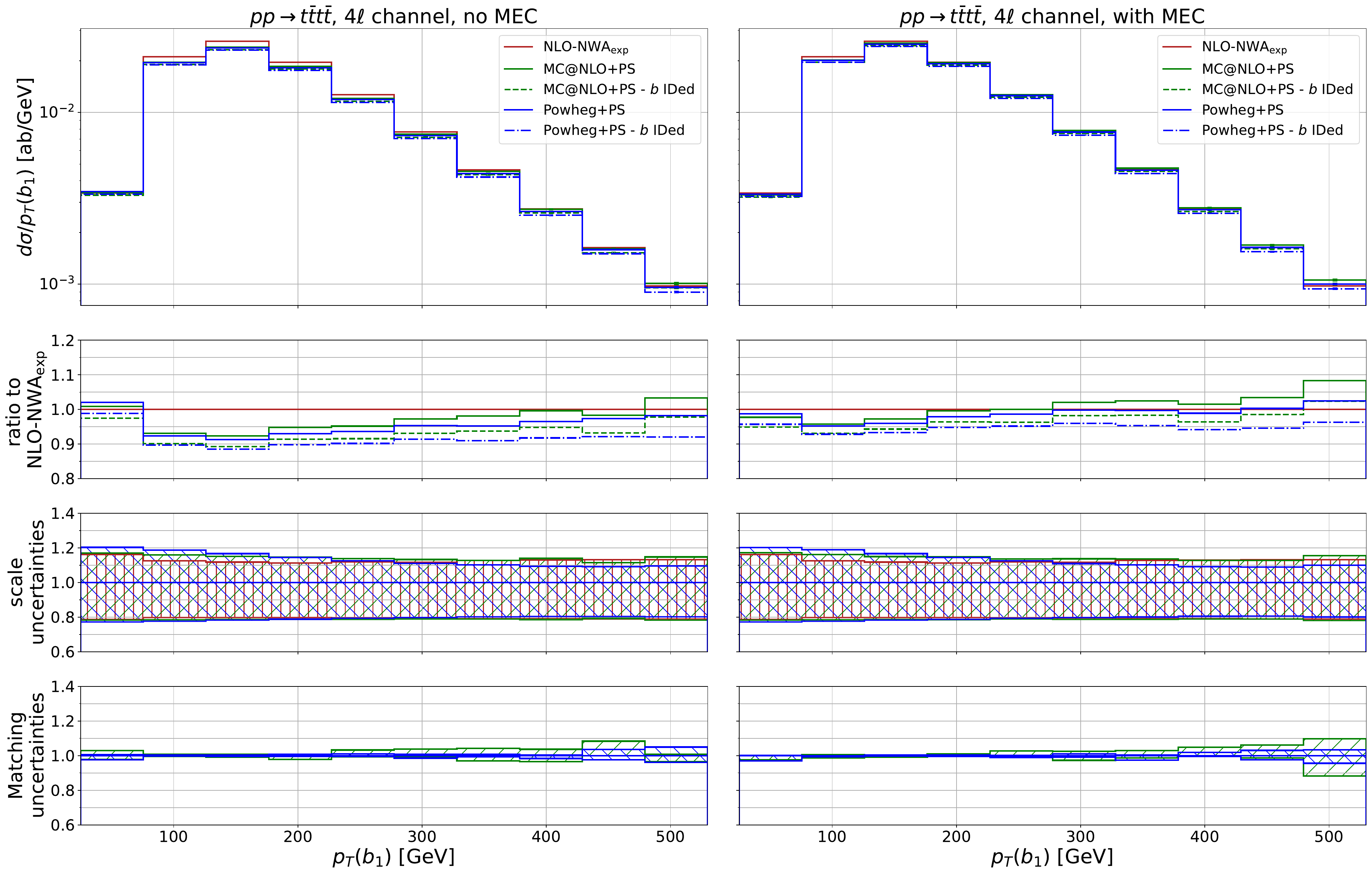}
    \\[0.4cm]
    \includegraphics[width=0.9\linewidth]{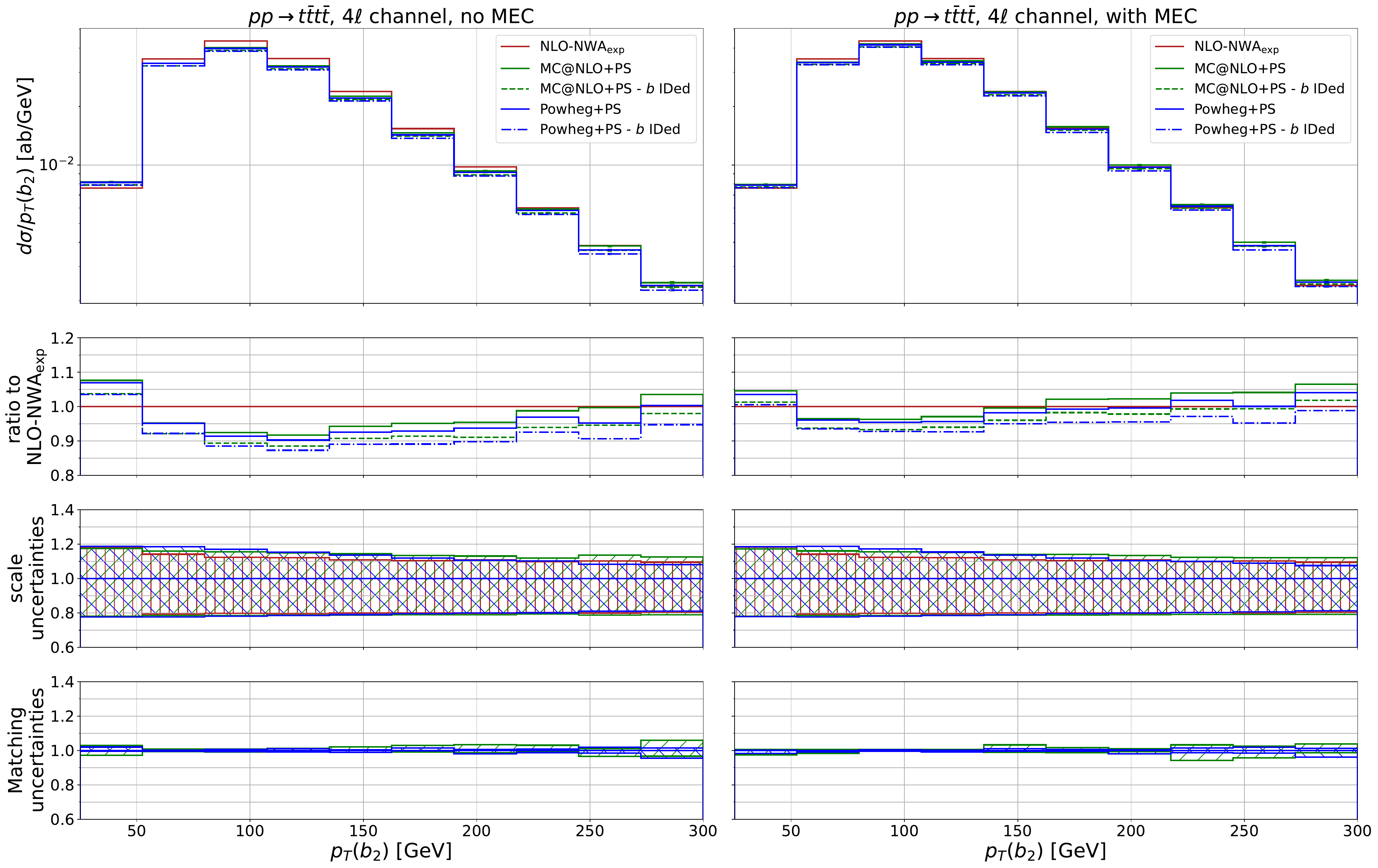}
    \caption{\textit{Differential cross-section distributions at NLO in QCD for the $pp\to t\bar{t}t\bar{t}+X$ process in the $4\ell$ channel as a function of the transverse momentum of the first and second hardest $b$-jets, $p_T(b_1)$ and $p_T(b_2)$, respectively. The parton-shower based results are given without (left) and with the MEC (right). Predictions without (solid line) and with (dashed line) the $b$-jet identification are also shown. The second panel from the top provides the ratios to $\sigma^{\rm NLO}_{\rm NWA_{exp}}$. The third panel depicts the scale uncertainties, and the bottom panel presents the  parton-shower matching uncertainties.}}
    \label{fig:pT_b1b2}
\end{figure}

\begin{figure}
    \centering
    \includegraphics[width=0.9\linewidth]{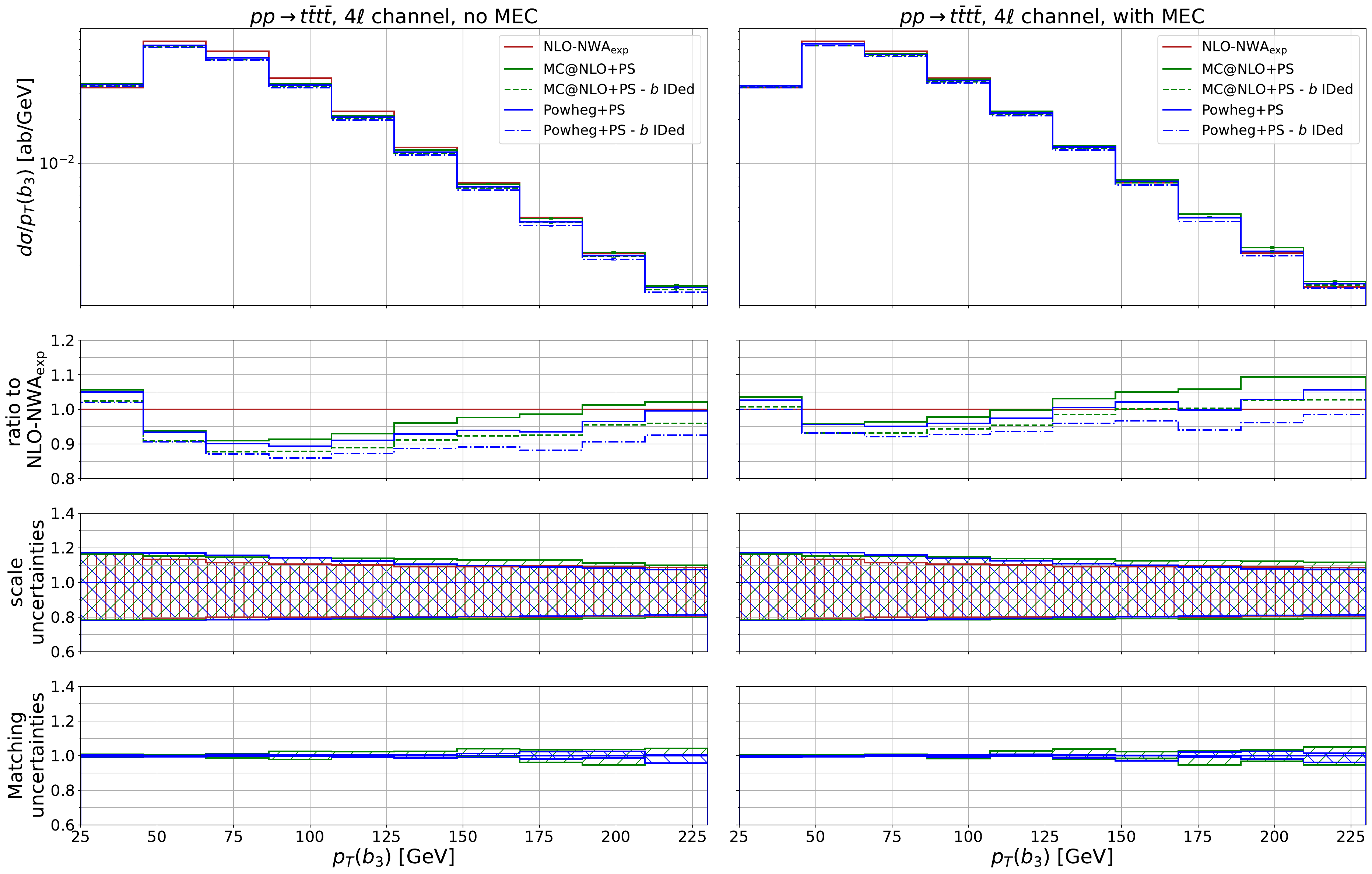}\\[0.4cm]
    \includegraphics[width=0.9\linewidth]{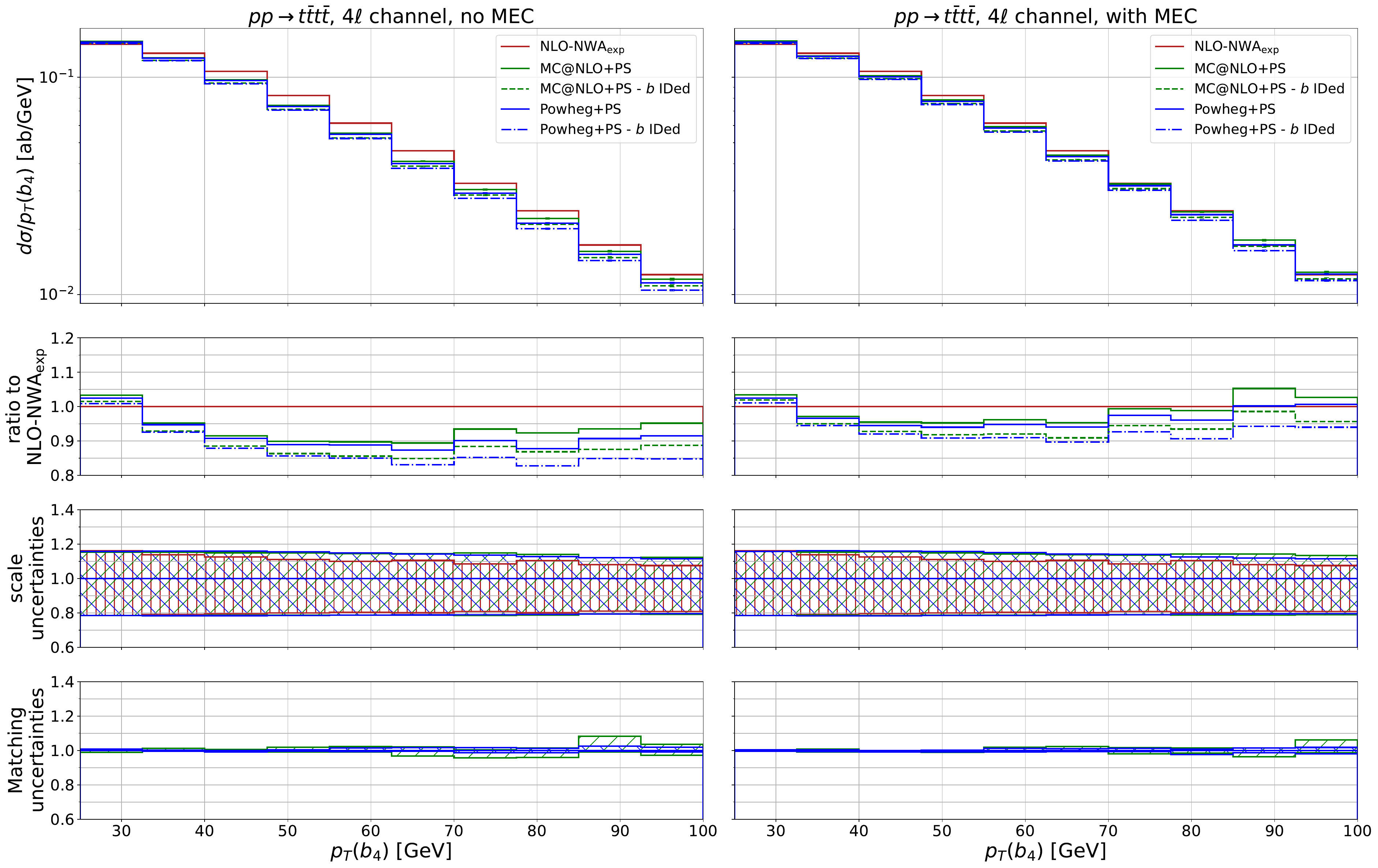}
    \caption{\textit{Same as Figure \ref{fig:pT_b1b2}
    but for the transverse momentum of the third and fourth hardest $b$-jets, $p_T(b_3)$ and $p_T(b_4)$, respectively.}}
    \label{fig:pT_b3b4}
\end{figure}

\begin{figure}[t]
    \centering
    \includegraphics[width=0.9\linewidth]{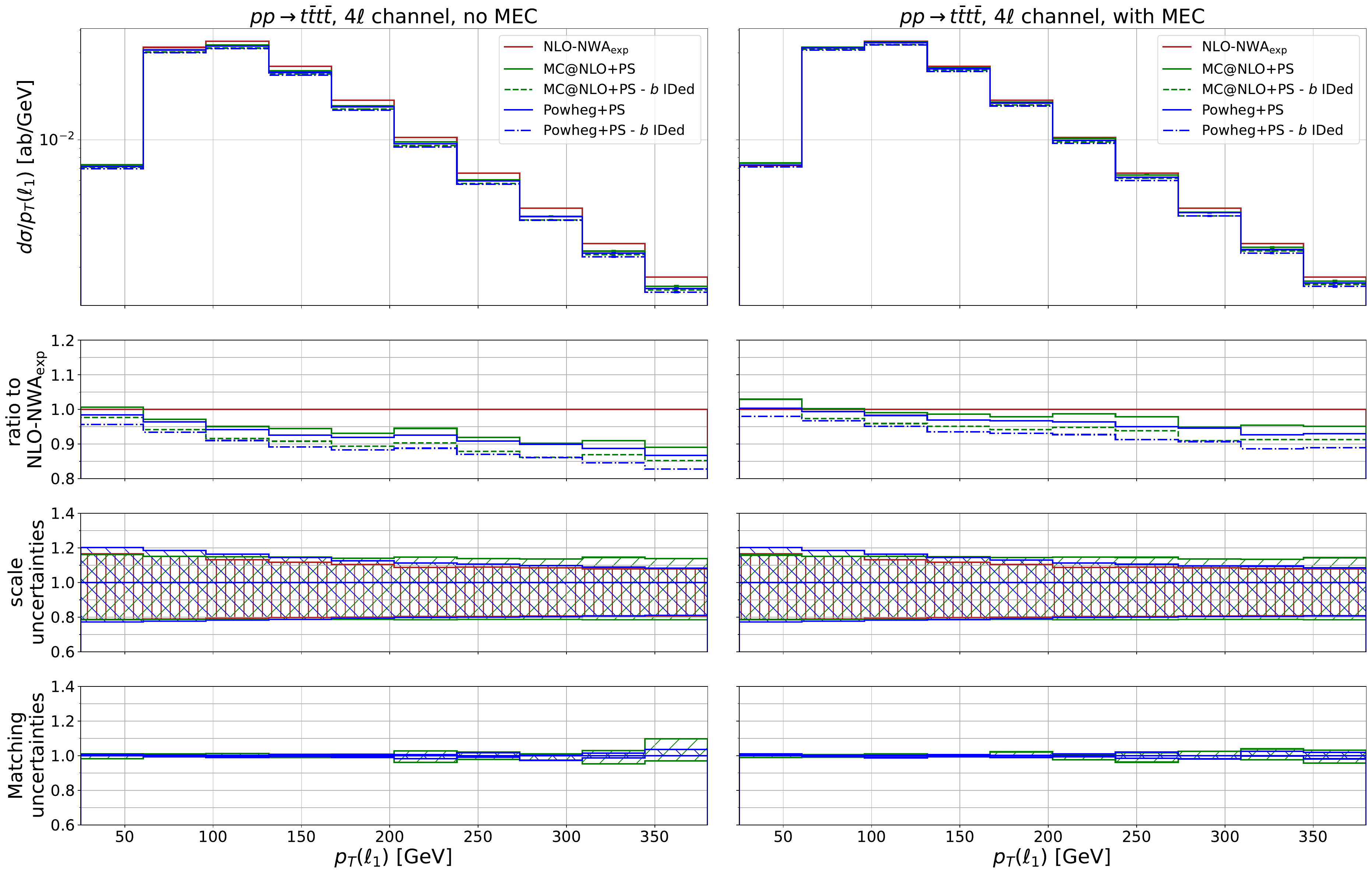}\\[0.4cm]
    \includegraphics[width=0.9\linewidth]{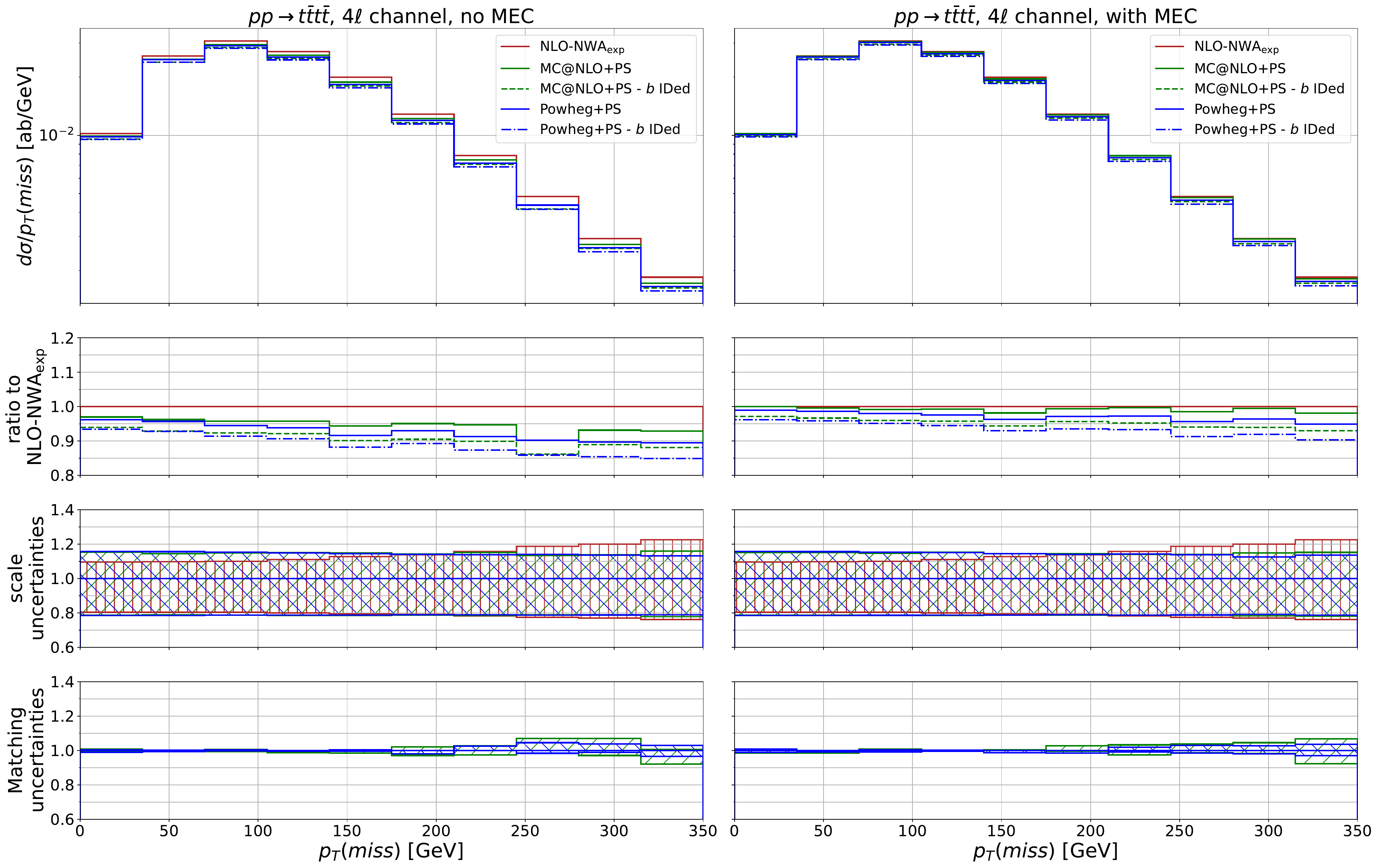}
    \caption{\textit{Same as Figure \ref{fig:pT_b1b2}
    but for the transverse momentum of the hardest charged lepton, $p_T(\ell_1)$ and for the missing transverse momentum, $p_T(miss)$.}}
    \label{fig:pT_l1_pT_miss}
\end{figure}

\begin{figure}
    \centering
    \includegraphics[width=0.9\linewidth]{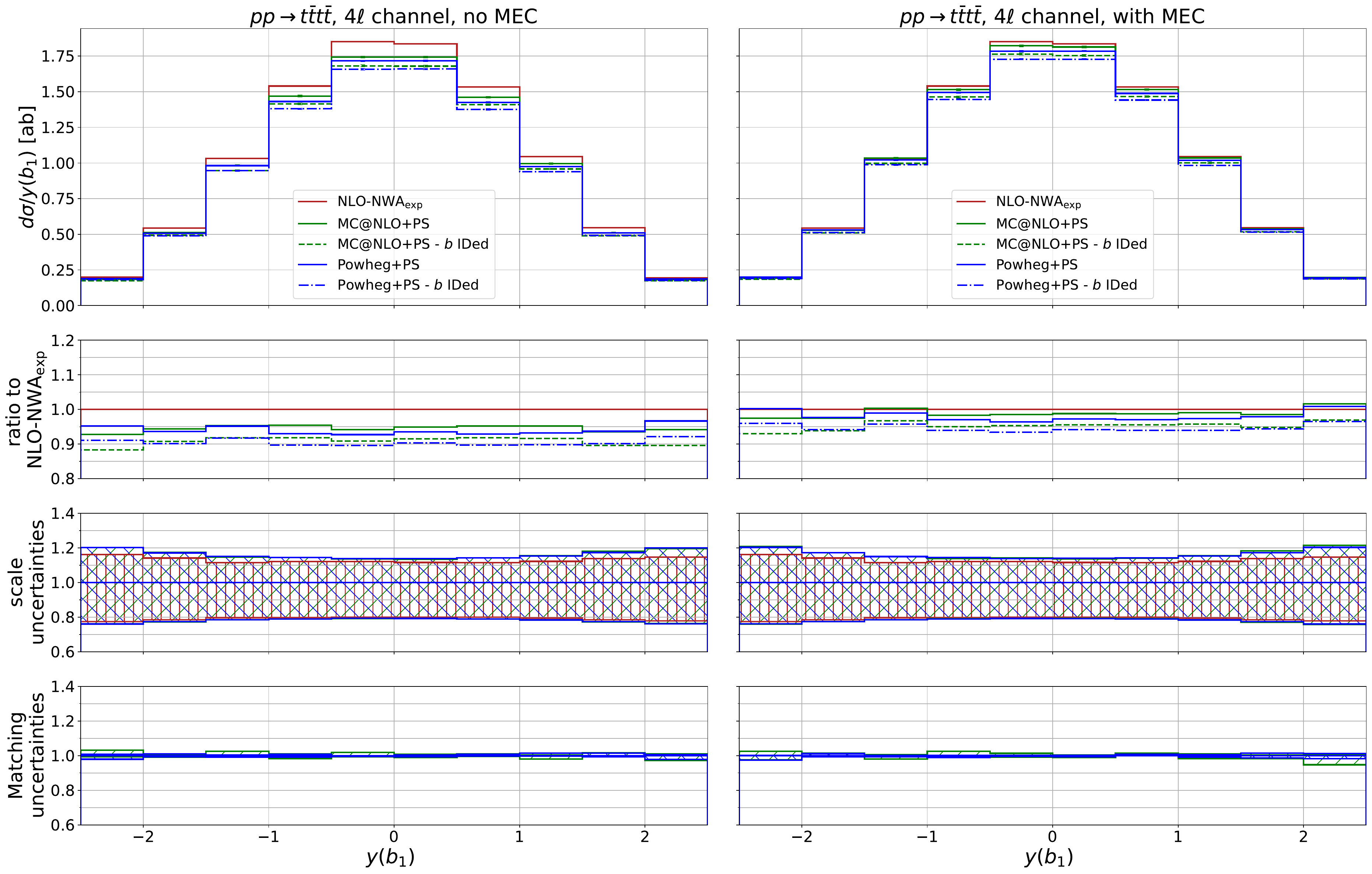}\\[0.4cm]
    \includegraphics[width=0.9\linewidth]{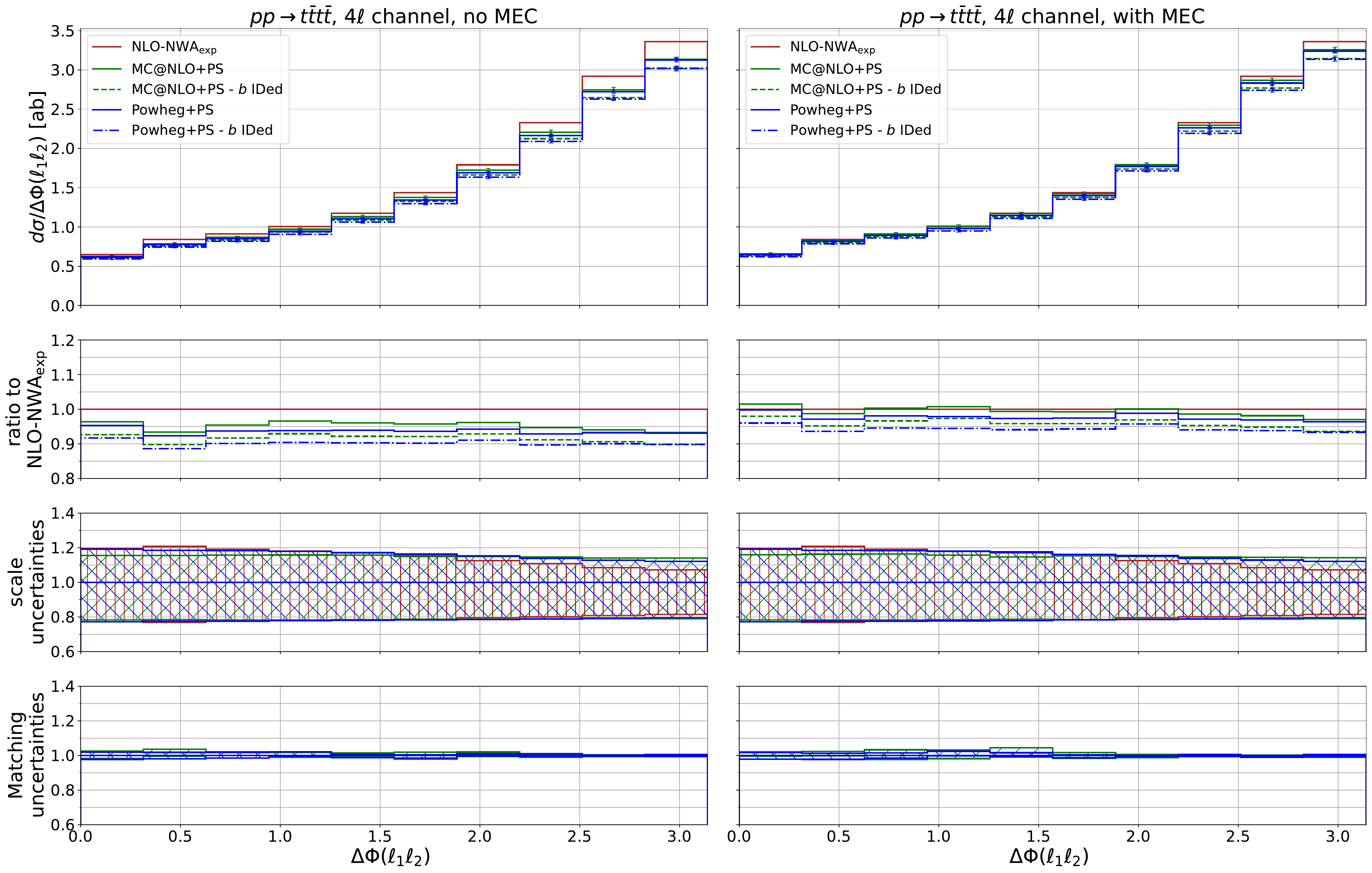}
    \caption{\textit{Same as Figure \ref{fig:pT_b1b2} but for the rapidity of the hardest $b$-jet, $y(b_1)$, and the angular distance between the first and second hardest (charged) leptons, $\Delta \phi(\ell_1\ell_2)$.}}
    \label{fig:y_b1_dPhi_l2l4}
\end{figure}

In the next step, we examine various differential cross-section distributions obtained by applying the cuts and input parameters specified in Section \ref{description}. We consider a set of both dimensionful and dimensionless observables. For dimensionful observables, the ranges are chosen to ensure the presence of at least a few events in the last bin, assuming the integrated luminosity of the HL-LHC  as well as including $\tau$ leptons in the definition of leptons. We remind the reader that the HL-LHC  will produce more than $250 \, {\rm fb}^{-1}$ of data per year and be capable of collecting up to $4000 \, {\rm fb}^{-1}$ of data  during its whole exploitation period. For each of the observables illustrated in Figures \ref{fig:pT_b1b2} - \ref{fig:y_b1_dPhi_l2l4} we present four plots and divide them into two classes: the results without and with MEC. Note that the results including MEC are only provided for the case of $\rm \oplus MEC$, since the differences between $\rm \oplus MEC^{1st}$ and $\rm \oplus MEC$ are negligible also at the differential cross-section level. For each plot, the upper panels show the absolute NLO predictions for the various approaches we considered. In particular, we show the fixed-order NLO QCD results in the NWA as well as the two parton-shower based predictions with and without the $b$-jet identification. In each case, the left plots correspond to the results without MEC, while the right plots include such effects. The second panel from the top provides the ratio to the $\sigma^{\rm NLO}_{\rm NWA_{exp}}$ result. The third panel depicts the uncertainty bands resulting from scale variations. In this case only the uncertainties for the predictions without the $b$-jet identification are plotted. However, we have verified that the identification of the $b$-jets does not change their size. Finally, the bottom panel presents the matching uncertainties for the parton-shower based results, again only for the predictions without the $b$-jet identification.  

In Figure \ref{fig:pT_b1b2} we show the transverse momentum of the first,  $p_T(b_1)$, and second hardest $b$-jets, $p_T(b_2)$. Prior to the application of the MEC, differences of up to $10\%$ are visible compared to the ${\sigma^{\rm NLO}_{\rm NWA_{exp}}}$ case. Much better consistency in the shape of these distributions can be observed after the MEC are included. Indeed, in the phase-space regions where most of the events are located, the various predictions now differ by only about $5\%$. In addition, the shape of the $p_T(b_1)$ and $p_T(b_2)$ distributions does not change significantly after the $b$-jet identification is performed. This indicates that the two hardest $b$-jets most likely come from the top-quark decays. Finally, also at the differential cross-section level, the two parton-shower based results agree with each other within the quoted theoretical uncertainties due to missing higher-order corrections. These uncertainties are of the order of $15\%-20\%$, thus, they are similar in size to those observed at the integrated cross-section level. In addition, the matching uncertainties are negligible. The latter effect is visible for all the observables we have examined for this process.

Next, we look at the transverse momentum distributions of the third, $p_T(b_3)$, and  fourth hardest $b$-jets, $p_T(b_4)$, that are presented in Figure \ref{fig:pT_b3b4}. In contrast to $p_T(b_1)$ and $p_T(b_2)$, for $p_T(b_3)$ and $p_T(b_4)$ we can observe more significant shape distortions when comparing the various predictions. If the MEC are not employed, shape changes of up to $10\%-15\%$ are visible when parton-shower based predictions are contrasted with the fixed-order ${\sigma^{\rm NLO}_{\rm NWA_{exp}}}$ result. These differences are reduced to less than $10\%$ if the MEC are enabled. The increasing deviations between the results without and with the $b$-jet identification indicate that the probability that the third and fourth hardest $b$-jets originate from the parton-shower simulations rather than from the top-quark decays increases significantly. One can also see larger discrepancies between the two parton-shower based predictions, especially in the case of the high $p_T$ tails of the two distributions. This is because the emissions in MC@NLO-like simulations, unlike  in \textsc{Powheg} simulations, are not ordered in hardness, i.e. the hardest emission is not necessarily generated before the parton shower is added. However, all presented results are consistent with each other due to relatively large theoretical uncertainties.

Turning to leptonic observables, in Figure \ref{fig:pT_l1_pT_miss} we plot the transverse momentum of the hardest (charged) lepton, $p_T(\ell_1)$, and 
the missing transverse momentum due to the four neutrinos, $p_T(miss)$. In the high $p_T$ tails, parton showers have a significant impact on both observables. We observe shower effects up to even $17\%$. The large discrepancy between the parton-shower matched results and the fixed-order predictions in these phase-space regions can be explained by the fact that multiple parton-shower emissions decrease the transverse momenta of the colour-neutral particles, which have to recoil against the sum of all emitted partons. Enabling the MEC brings the different predictions closer together. Indeed, the deviations are now reduced to about $10\%$.

Finally, we analyse two dimensionless distributions. In Figure \ref{fig:y_b1_dPhi_l2l4} we plot the rapidity of the hardest $b$-jet, $y(b_1)$, and the angular distance between the first and second hardest charged leptons, $\Delta \phi (\ell_1\ell_2)$. The $\Delta \phi (\ell_1\ell_2)$ observable is sensitive to new physical scenarios in which new heavy states can be produced that further decay into top quarks. In general, the angular distributions of charged leptons are of great importance, because they reflect spin correlations  and can be used to probe the ${\cal CP}$ quantum numbers of such new states. For both observables, we note that the ratio of the parton-shower based predictions to fixed-order results is flat, with any differences arising solely from changes in normalisation. We verified this explicitly by also generating the normalised differential cross-section distributions.  This  normalisation change is in the range of $5\%-10\%$ depending on whether an attempt is made to identify the $b$-jets from top-quark decays. Also for these two observables, the MEC method helps to bring the different predictions closer together. The difference in the normalisation factor is reduced to a value below $5\%$, which is within the obtained theoretical uncertainties. Although we have only shown the rapidity of the hardest $b$-jet here, the same conclusions can be drawn for the remaining three $b$-jets, i.e. for $y(b_2)$, $y(b_3)$ and $y(b_4)$. Similarly, every other combination of leptons that we used to construct $\Delta \phi (\ell\ell)$ behaved similarly to the $\Delta \phi (\ell_1\ell_2)$ differential cross-section distribution for all the theoretical predictions that we considered.

%
\section{Summary and Conclusions}
\label{sec:outlook}
%

In this paper we presented a comparison of the fixed-order NLO QCD results for the $pp\to t\bar{t}t\bar{t}+X$ process in the $4\ell$ channel, with the parton-shower based predictions obtained using  the MC@NLO and \textsc{Powheg} matching. In the fixed-order case, all higher-order QCD effects were consistently incorporated into the production and decays of the four top quarks. This calculation was performed in the NWA, which preserves the NLO spin correlations throughout the computational process. In the second case, NLO QCD corrections were only included in the production stage of the $pp\to t\bar{t}t\bar{t}+X$ process, whereas top-quark decays were treated at LO, thus, neglecting NLO spin correlations. Furthermore, we examined the importance of using the MEC in top-quark decays for the parton-shower based predictions. Such a multilayer comparison helped us to assess to what extent parton showers could approximate the higher-order QCD effects in top-quark decays. We investigated the validity of these approximations in such a complex environment, both at the integrated and differential cross-section level. Finally, the presented comparison allowed us to identify observables and specific phase-space regions that are sensitive to the resummed dominant soft-collinear logarithmic corrections from parton showers, which are not present in the fixed-order NLO QCD predictions.

To conclude, at the integrated cross-section level all the parton-shower based results were smaller than the predictions obtained at the fixed-order level. This is due to the fact that multiple shower emissions allow the $b$-jets originating from the matrix-element level to go below the given kinematic cut. Moreover, neglecting completely NLO QCD effects in top-quark decays increased the cross section by about $9\%$ and the theoretical uncertainties by $3\%$. The \textsc{Powheg} and MC@NLO results recovered some of these higher-order effects. Indeed, the NLO plus parton-shower predictions differed from the $\sigma^{\rm NLO}_{\rm NWA_{exp}}$ result by $5\%-7\%$. After including the MEC, these differences were reduced to $1\%-3\%$, highlighting the importance of enabling such effects in the parton-shower based predictions. When trying to identify the $b$-jets from top-quark decays, we observed that about $3\%-5\%$ more events were vetoed. As a result, the \textsc{Powheg} and MC@NLO predictions differed from the fixed-order result by  $10\%$. These differences were also reduced to $5\%-6\%$ after employing the MEC. We also investigated the case where the use of the MEC was not allowed after the first parton-shower emission. As expected, the difference between the two results $\oplus{\rm MEC}$ and $\oplus{\rm MEC^{1st}}$ was very small, less than $1\%$. Although it was very interesting and important to understand these various differences, we must add here that at present they all fall within the estimated theoretical uncertainties, which are of the order of $20\%$.

In the case of differential cross-section distributions for all the observables considered, the size of the theoretical uncertainties was driven by the size of the scale uncertainties. Also at the differential cross-section level these  uncertainties were consistently of the order of $20\%$. We observed that the parton-shower matching uncertainties were negligible for all the observables we analysed. Furthermore, we could clearly notice that enabling the MEC algorithm in \textsc{Pythia} improved the modelling of top-quark decays in the shower evolution for all the observables we examined. Moreover, we saw good consistency between the results obtained with the \textsc{Powheg} and MC@NLO matching. We also examined how various differential cross-section distributions were modified by the showering stage and in particular saw that there was a significant (indirect) impact on the leptonic observables. On the contrary, similar large effects were not observed when looking, for example, at the two hardest $b$-jets. On average, the shower emissions were softer in $p_T$ with respect to the first and second hardest $b$-jet from the top-quark decays. In these cases, in the high $p_T$ tails the obtained predictions were closer to the fixed-order NLO QCD results. This  was especially visible in the case of the results obtained within the \textsc{Powheg} framework. Finally, another very important topic that we investigated was spin correlations. Having fixed-order predictions describing top-quark decays and spin correlations with NLO QCD accuracy, we were in an excellent position to test the impact of parton showers and MEC on the angular distributions. Overall, we observed that for these distributions the differences between the parton-shower matched results and the fixed-order predictions were in the range of $5\%-10\%$. Introducing the MEC helped to reduce them to less than $5\%$.

\subsection*{ACKNOWLEDGMENTS}

We would like to thank Manfred Kraus for very helpful discussions regarding the results presented in Ref. \cite{Jezo:2021smh}.

This work was supported by the Deutsche Forschungsgemeinschaft (DFG) under grant 396021762 - TRR 257: \textit{Particle Physics Phenomenology after the Higgs Discovery}, and grant 400140256 - GRK 2497: \textit{The Physics of the Heaviest Particles at the LHC.} Support by a grant of the Bundesministerium f\"ur Bildung und Forschung (BMBF) is additionally acknowledged.

The authors gratefully acknowledge the computing time provided to them at the NHR Center NHR4CES at RWTH Aachen University (project number \texttt{p0020216}). This is funded by the Federal Ministry of Education and Research, and the state governments participating on the basis of the
resolutions of the GWK for national high performance computing at universities.

\bibliographystyle{utphys}

\begin{thebibliography}{10}

\bibitem{ATLAS:2023ajo}
{\bfseries ATLAS} Collaboration, G.~Aad {\em et~al.}, ``{Observation of
  four-top-quark production in the multilepton final state with the ATLAS
  detector},'' \href{http://dx.doi.org/10.1140/epjc/s10052-023-11573-0}{{\em
  Eur. Phys. J. C} {\bfseries 83} no.~6, (2023) 496},
  \href{http://arxiv.org/abs/2303.15061}{{\ttfamily arXiv:2303.15061
  [hep-ex]}}. [Erratum: Eur.Phys.J.C 84, 156 (2024)].

\bibitem{CMS:2023ftu}
{\bfseries CMS} Collaboration, A.~Hayrapetyan {\em et~al.}, ``{Observation of
  four top quark production in proton-proton collisions at $\sqrt{s}=13$
  TeV},'' \href{http://dx.doi.org/10.1016/j.physletb.2023.138290}{{\em Phys.
  Lett. B} {\bfseries 847} (2023) 138290},
  \href{http://arxiv.org/abs/2305.13439}{{\ttfamily arXiv:2305.13439
  [hep-ex]}}.

\bibitem{Bevilacqua:2012em}
G.~Bevilacqua and M.~Worek, ``{Constraining BSM Physics at the LHC: Four top
  final states with NLO accuracy in perturbative QCD},''
  \href{http://dx.doi.org/10.1007/JHEP07(2012)111}{{\em JHEP} {\bfseries 07}
  (2012) 111}, \href{http://arxiv.org/abs/1206.3064}{{\ttfamily arXiv:1206.3064
  [hep-ph]}}.

\bibitem{Alwall:2014hca}
J.~Alwall, R.~Frederix, S.~Frixione, V.~Hirschi, F.~Maltoni, O.~Mattelaer,
  H.~S. Shao, T.~Stelzer, P.~Torrielli, and M.~Zaro, ``{The automated
  computation of tree-level and next-to-leading order differential cross
  sections, and their matching to parton shower simulations},''
  \href{http://dx.doi.org/10.1007/JHEP07(2014)079}{{\em JHEP} {\bfseries 07}
  (2014) 079}, \href{http://arxiv.org/abs/1405.0301}{{\ttfamily arXiv:1405.0301
  [hep-ph]}}.

\bibitem{Maltoni:2015ena}
F.~Maltoni, D.~Pagani, and I.~Tsinikos, ``{Associated production of a top-quark
  pair with vector bosons at NLO in QCD: impact on $
  \mathrm{t}\overline{\mathrm{t}}\mathrm{H} $ searches at the LHC},''
  \href{http://dx.doi.org/10.1007/JHEP02(2016)113}{{\em JHEP} {\bfseries 02}
  (2016) 113}, \href{http://arxiv.org/abs/1507.05640}{{\ttfamily
  arXiv:1507.05640 [hep-ph]}}.

\bibitem{Frederix:2017wme}
R.~Frederix, D.~Pagani, and M.~Zaro, ``{Large NLO corrections in
  $t\bar{t}W^{\pm}$ and $t\bar{t}t\bar{t}$ hadroproduction from supposedly
  subleading EW contributions},''
  \href{http://dx.doi.org/10.1007/JHEP02(2018)031}{{\em JHEP} {\bfseries 02}
  (2018) 031}, \href{http://arxiv.org/abs/1711.02116}{{\ttfamily
  arXiv:1711.02116 [hep-ph]}}.

\bibitem{vanBeekveld:2022hty}
M.~van Beekveld, A.~Kulesza, and L.~M. Valero, ``{Threshold Resummation for the
  Production of Four Top Quarks at the LHC},''
  \href{http://dx.doi.org/10.1103/PhysRevLett.131.211901}{{\em Phys. Rev.
  Lett.} {\bfseries 131} no.~21, (2023) 211901},
  \href{http://arxiv.org/abs/2212.03259}{{\ttfamily arXiv:2212.03259
  [hep-ph]}}.

\bibitem{vanBeekveld:2025ghw}
M.~van Beekveld, A.~Kulesza, M.~Lupattelli, and T.~Saracco, ``{Invariant-mass
  threshold resummation for the production of four top quarks at the LHC},''
  \href{http://arxiv.org/abs/2505.10381}{{\ttfamily arXiv:2505.10381
  [hep-ph]}}.

\bibitem{Jezo:2021smh}
T.~Je\v{z}o and M.~Kraus, ``{Hadroproduction of four top quarks in the powheg
  box},'' \href{http://dx.doi.org/10.1103/PhysRevD.105.114024}{{\em Phys. Rev.
  D} {\bfseries 105} no.~11, (2022) 114024},
  \href{http://arxiv.org/abs/2110.15159}{{\ttfamily arXiv:2110.15159
  [hep-ph]}}.

\bibitem{Dimitrakopoulos:2024qib}
N.~Dimitrakopoulos and M.~Worek, ``{Four top final states with NLO accuracy in
  perturbative QCD: 4 lepton channel},''
  \href{http://dx.doi.org/10.1007/JHEP06(2024)129}{{\em JHEP} {\bfseries 06}
  (2024) 129}, \href{http://arxiv.org/abs/2401.10678}{{\ttfamily
  arXiv:2401.10678 [hep-ph]}}.

\bibitem{Dimitrakopoulos:2024yjm}
N.~Dimitrakopoulos and M.~Worek, ``{Four top final states with NLO accuracy in
  perturbative QCD: 3 lepton channel},''
  \href{http://dx.doi.org/10.1007/JHEP03(2025)025}{{\em JHEP} {\bfseries 03}
  (2025) 025}, \href{http://arxiv.org/abs/2410.05960}{{\ttfamily
  arXiv:2410.05960 [hep-ph]}}.

\bibitem{Bailey:2020ooq}
S.~Bailey, T.~Cridge, L.~A. Harland-Lang, A.~D. Martin, and R.~S. Thorne,
  ``{Parton distributions from LHC, HERA, Tevatron and fixed target data:
  MSHT20 PDFs},'' \href{http://dx.doi.org/10.1140/epjc/s10052-021-09057-0}{{\em
  Eur. Phys. J. C} {\bfseries 81} no.~4, (2021) 341},
  \href{http://arxiv.org/abs/2012.04684}{{\ttfamily arXiv:2012.04684
  [hep-ph]}}.

\bibitem{Buckley:2014ana}
A.~Buckley, J.~Ferrando, S.~Lloyd, K.~Nordstr\"om, B.~Page, M.~R\"ufenacht,
  M.~Sch\"onherr, and G.~Watt, ``{LHAPDF6: parton density access in the LHC
  precision era},''
  \href{http://dx.doi.org/10.1140/epjc/s10052-015-3318-8}{{\em Eur. Phys. J. C}
  {\bfseries 75} (2015) 132}, \href{http://arxiv.org/abs/1412.7420}{{\ttfamily
  arXiv:1412.7420 [hep-ph]}}.

\bibitem{Cacciari:2008gp}
M.~Cacciari, G.~P. Salam, and G.~Soyez, ``{The anti-$k_t$ jet clustering
  algorithm},'' \href{http://dx.doi.org/10.1088/1126-6708/2008/04/063}{{\em
  JHEP} {\bfseries 04} (2008) 063},
  \href{http://arxiv.org/abs/0802.1189}{{\ttfamily arXiv:0802.1189 [hep-ph]}}.

\bibitem{Bevilacqua:2011xh}
G.~Bevilacqua, M.~Czakon, M.~V. Garzelli, A.~van Hameren, A.~Kardos, C.~G.
  Papadopoulos, R.~Pittau, and M.~Worek, ``{HELAC-NLO},''
  \href{http://dx.doi.org/10.1016/j.cpc.2012.10.033}{{\em Comput. Phys.
  Commun.} {\bfseries 184} (2013) 986--997},
  \href{http://arxiv.org/abs/1110.1499}{{\ttfamily arXiv:1110.1499 [hep-ph]}}.

\bibitem{Ossola:2008xq}
G.~Ossola, C.~G. Papadopoulos, and R.~Pittau, ``{On the Rational Terms of the
  one-loop amplitudes},''
  \href{http://dx.doi.org/10.1088/1126-6708/2008/05/004}{{\em JHEP} {\bfseries
  05} (2008) 004}, \href{http://arxiv.org/abs/0802.1876}{{\ttfamily
  arXiv:0802.1876 [hep-ph]}}.

\bibitem{vanHameren:2009dr}
A.~van Hameren, C.~G. Papadopoulos, and R.~Pittau, ``{Automated one-loop
  calculations: A Proof of concept},''
  \href{http://dx.doi.org/10.1088/1126-6708/2009/09/106}{{\em JHEP} {\bfseries
  09} (2009) 106}, \href{http://arxiv.org/abs/0903.4665}{{\ttfamily
  arXiv:0903.4665 [hep-ph]}}.

\bibitem{vanHameren:2010cp}
A.~van Hameren, ``{OneLOop: For the evaluation of one-loop scalar functions},''
  \href{http://dx.doi.org/10.1016/j.cpc.2011.06.011}{{\em Comput. Phys.
  Commun.} {\bfseries 182} (2011) 2427--2438},
  \href{http://arxiv.org/abs/1007.4716}{{\ttfamily arXiv:1007.4716 [hep-ph]}}.

\bibitem{Czakon:2009ss}
M.~Czakon, C.~G. Papadopoulos, and M.~Worek, ``{Polarizing the Dipoles},''
  \href{http://dx.doi.org/10.1088/1126-6708/2009/08/085}{{\em JHEP} {\bfseries
  08} (2009) 085}, \href{http://arxiv.org/abs/0905.0883}{{\ttfamily
  arXiv:0905.0883 [hep-ph]}}.

\bibitem{Bevilacqua:2009zn}
G.~Bevilacqua, M.~Czakon, C.~G. Papadopoulos, R.~Pittau, and M.~Worek,
  ``{Assault on the NLO Wishlist: $pp\to t\bar{t}b\bar{b}$},''
  \href{http://dx.doi.org/10.1088/1126-6708/2009/09/109}{{\em JHEP} {\bfseries
  09} (2009) 109}, \href{http://arxiv.org/abs/0907.4723}{{\ttfamily
  arXiv:0907.4723 [hep-ph]}}.

\bibitem{Bevilacqua:2013iha}
G.~Bevilacqua, M.~Czakon, M.~Kubocz, and M.~Worek, ``{Complete Nagy-Soper
  subtraction for next-to-leading order calculations in QCD},''
  \href{http://dx.doi.org/10.1007/JHEP10(2013)204}{{\em JHEP} {\bfseries 10}
  (2013) 204}, \href{http://arxiv.org/abs/1308.5605}{{\ttfamily arXiv:1308.5605
  [hep-ph]}}.

\bibitem{Alwall:2006yp}
J.~Alwall {\em et~al.}, ``{A Standard format for Les Houches event files},''
  \href{http://dx.doi.org/10.1016/j.cpc.2006.11.010}{{\em Comput. Phys.
  Commun.} {\bfseries 176} (2007) 300--304},
  \href{http://arxiv.org/abs/hep-ph/0609017}{{\ttfamily arXiv:hep-ph/0609017}}.

\bibitem{Antcheva:2009zz}
I.~Antcheva {\em et~al.}, ``{ROOT: A C++ framework for petabyte data storage,
  statistical analysis and visualization},''
  \href{http://dx.doi.org/10.1016/j.cpc.2009.08.005}{{\em Comput. Phys.
  Commun.} {\bfseries 180} (2009) 2499--2512},
  \href{http://arxiv.org/abs/1508.07749}{{\ttfamily arXiv:1508.07749
  [physics.data-an]}}.

\bibitem{Bern:2013zja}
Z.~Bern, L.~J. Dixon, F.~Febres~Cordero, S.~H\"oche, H.~Ita, D.~A. Kosower, and
  D.~Maitre, ``{Ntuples for NLO Events at Hadron Colliders},''
  \href{http://dx.doi.org/10.1016/j.cpc.2014.01.011}{{\em Comput. Phys.
  Commun.} {\bfseries 185} (2014) 1443--1460},
  \href{http://arxiv.org/abs/1310.7439}{{\ttfamily arXiv:1310.7439 [hep-ph]}}.

\bibitem{Bevilacqua:HEPlot}
G.~Bevilacqua, ``unpublished.'' 2019.

\bibitem{Nason:2004rx}
P.~Nason, ``{A New method for combining NLO QCD with shower Monte Carlo
  algorithms},'' \href{http://dx.doi.org/10.1088/1126-6708/2004/11/040}{{\em
  JHEP} {\bfseries 11} (2004) 040},
  \href{http://arxiv.org/abs/hep-ph/0409146}{{\ttfamily arXiv:hep-ph/0409146}}.

\bibitem{Frixione:2007vw}
S.~Frixione, P.~Nason, and C.~Oleari, ``{Matching NLO QCD computations with
  Parton Shower simulations: the POWHEG method},''
  \href{http://dx.doi.org/10.1088/1126-6708/2007/11/070}{{\em JHEP} {\bfseries
  11} (2007) 070}, \href{http://arxiv.org/abs/0709.2092}{{\ttfamily
  arXiv:0709.2092 [hep-ph]}}.

\bibitem{Alioli:2010xd}
S.~Alioli, P.~Nason, C.~Oleari, and E.~Re, ``{A general framework for
  implementing NLO calculations in shower Monte Carlo programs: the POWHEG
  BOX},'' \href{http://dx.doi.org/10.1007/JHEP06(2010)043}{{\em JHEP}
  {\bfseries 06} (2010) 043}, \href{http://arxiv.org/abs/1002.2581}{{\ttfamily
  arXiv:1002.2581 [hep-ph]}}.

\bibitem{Frixione:2002ik}
S.~Frixione and B.~R. Webber, ``{Matching NLO QCD computations and parton
  shower simulations},''
  \href{http://dx.doi.org/10.1088/1126-6708/2002/06/029}{{\em JHEP} {\bfseries
  06} (2002) 029}, \href{http://arxiv.org/abs/hep-ph/0204244}{{\ttfamily
  arXiv:hep-ph/0204244}}.

\bibitem{Bierlich:2022pfr}
C.~Bierlich {\em et~al.}, ``{A comprehensive guide to the physics and usage of
  PYTHIA 8.3}'' \href{http://dx.doi.org/10.21468/SciPostPhysCodeb.8}{{\em
  SciPost Phys. Codeb.} {\bfseries 2022} (2022) 8},
  \href{http://arxiv.org/abs/2203.11601}{{\ttfamily arXiv:2203.11601
  [hep-ph]}}.

\bibitem{Buckley:2010ar}
A.~Buckley, J.~Butterworth, D.~Grellscheid, H.~Hoeth, L.~Lonnblad, J.~Monk,
  H.~Schulz, and F.~Siegert, ``{Rivet user manual},''
  \href{http://dx.doi.org/10.1016/j.cpc.2013.05.021}{{\em Comput. Phys.
  Commun.} {\bfseries 184} (2013) 2803--2819},
  \href{http://arxiv.org/abs/1003.0694}{{\ttfamily arXiv:1003.0694 [hep-ph]}}.

\bibitem{Bierlich:2019rhm}
C.~Bierlich {\em et~al.}, ``{Robust Independent Validation of Experiment and
  Theory: Rivet version 3},''
  \href{http://dx.doi.org/10.21468/SciPostPhys.8.2.026}{{\em SciPost Phys.}
  {\bfseries 8} (2020) 026}, \href{http://arxiv.org/abs/1912.05451}{{\ttfamily
  arXiv:1912.05451 [hep-ph]}}.

\bibitem{Buckley:2019xhk}
A.~Buckley, P.~Ilten, D.~Konstantinov, L.~L\"onnblad, J.~Monk, W.~Pokorski,
  T.~Przedzinski, and A.~Verbytskyi, ``{The HepMC3 event record library for
  Monte Carlo event generators},''
  \href{http://dx.doi.org/10.1016/j.cpc.2020.107310}{{\em Comput. Phys.
  Commun.} {\bfseries 260} (2021) 107310},
  \href{http://arxiv.org/abs/1912.08005}{{\ttfamily arXiv:1912.08005
  [hep-ph]}}.

\bibitem{Cacciari:2005hq}
M.~Cacciari and G.~P. Salam, ``{Dispelling the $N^{3}$ myth for the $k_t$
  jet-finder},'' \href{http://dx.doi.org/10.1016/j.physletb.2006.08.037}{{\em
  Phys. Lett. B} {\bfseries 641} (2006) 57--61},
  \href{http://arxiv.org/abs/hep-ph/0512210}{{\ttfamily arXiv:hep-ph/0512210}}.

\bibitem{Cacciari:2011ma}
M.~Cacciari, G.~P. Salam, and G.~Soyez, ``{FastJet User Manual},''
  \href{http://dx.doi.org/10.1140/epjc/s10052-012-1896-2}{{\em Eur. Phys. J. C}
  {\bfseries 72} (2012) 1896}, \href{http://arxiv.org/abs/1111.6097}{{\ttfamily
  arXiv:1111.6097 [hep-ph]}}.

\bibitem{Frixione:2023hwz}
S.~Frixione, S.~Amoroso, and S.~Mrenna, ``{Matrix element corrections in the
  Pythia8 parton shower in the context of matched simulations at
  next-to-leading order},''
  \href{http://dx.doi.org/10.1140/epjc/s10052-023-12154-x}{{\em Eur. Phys. J.
  C} {\bfseries 83} no.~10, (2023) 970},
  \href{http://arxiv.org/abs/2308.06389}{{\ttfamily arXiv:2308.06389
  [hep-ph]}}.

\bibitem{Frederix:2024psm}
R.~Frederix, L.~Gellersen, and J.~Nasufi, ``{Matrix element corrections in top
  quark decays for the $t\overline{t}W^{\pm }$ process},''
  \href{http://dx.doi.org/10.1140/epjc/s10052-024-12737-2}{{\em Eur. Phys. J.
  C} {\bfseries 84} no.~4, (2024) 410},
  \href{http://arxiv.org/abs/2402.12893}{{\ttfamily arXiv:2402.12893
  [hep-ph]}}.

\bibitem{Artoisenet:2012st}
P.~Artoisenet, R.~Frederix, O.~Mattelaer, and R.~Rietkerk, ``{Automatic
  spin-entangled decays of heavy resonances in Monte Carlo simulations},''
  \href{http://dx.doi.org/10.1007/JHEP03(2013)015}{{\em JHEP} {\bfseries 03}
  (2013) 015}, \href{http://arxiv.org/abs/1212.3460}{{\ttfamily arXiv:1212.3460
  [hep-ph]}}.

\bibitem{Frixione:2007zp}
S.~Frixione, E.~Laenen, P.~Motylinski, and B.~R. Webber, ``{Angular
  correlations of lepton pairs from vector boson and top quark decays in Monte
  Carlo simulations},''
  \href{http://dx.doi.org/10.1088/1126-6708/2007/04/081}{{\em JHEP} {\bfseries
  04} (2007) 081}, \href{http://arxiv.org/abs/hep-ph/0702198}{{\ttfamily
  arXiv:hep-ph/0702198}}.

\bibitem{FebresCordero:2021kcc}
F.~Febres~Cordero, M.~Kraus, and L.~Reina, ``{Top-quark pair production in
  association with a $W^\pm$ gauge boson in the POWHEG-BOX},''
  \href{http://dx.doi.org/10.1103/PhysRevD.103.094014}{{\em Phys. Rev. D}
  {\bfseries 103} no.~9, (2021) 094014},
  \href{http://arxiv.org/abs/2101.11808}{{\ttfamily arXiv:2101.11808
  [hep-ph]}}.

\bibitem{Bevilacqua:2021cit}
G.~Bevilacqua, H.-Y. Bi, H.~B. Hartanto, M.~Kraus, M.~Lupattelli, and M.~Worek,
  ``{$ t\overline{t}b\overline{b} $ at the LHC: on the size of corrections and
  b-jet definitions},'' \href{http://dx.doi.org/10.1007/JHEP08(2021)008}{{\em
  JHEP} {\bfseries 08} (2021) 008},
  \href{http://arxiv.org/abs/2105.08404}{{\ttfamily arXiv:2105.08404
  [hep-ph]}}.

\bibitem{Czakon:2022wam}
M.~Czakon, A.~Mitov, and R.~Poncelet, ``{Infrared-safe flavoured anti-k$_{T}$
  jets},'' \href{http://dx.doi.org/10.1007/JHEP04(2023)138}{{\em JHEP}
  {\bfseries 04} (2023) 138}, \href{http://arxiv.org/abs/2205.11879}{{\ttfamily
  arXiv:2205.11879 [hep-ph]}}.

\bibitem{Gauld:2022lem}
R.~Gauld, A.~Huss, and G.~Stagnitto, ``{Flavor Identification of Reconstructed
  Hadronic Jets},''
  \href{http://dx.doi.org/10.1103/PhysRevLett.130.161901}{{\em Phys. Rev.
  Lett.} {\bfseries 130} no.~16, (2023) 161901},
  \href{http://arxiv.org/abs/2208.11138}{{\ttfamily arXiv:2208.11138
  [hep-ph]}}. [Erratum: Phys.Rev.Lett. 132, 159901 (2024)].

\bibitem{Caola:2023wpj}
F.~Caola, R.~Grabarczyk, M.~L. Hutt, G.~P. Salam, L.~Scyboz, and J.~Thaler,
  ``{Flavored jets with exact anti-kt kinematics and tests of infrared and
  collinear safety},''
  \href{http://dx.doi.org/10.1103/PhysRevD.108.094010}{{\em Phys. Rev. D}
  {\bfseries 108} no.~9, (2023) 094010},
  \href{http://arxiv.org/abs/2306.07314}{{\ttfamily arXiv:2306.07314
  [hep-ph]}}.

\bibitem{Behring:2025ilo}
A.~Behring {\em et~al.}, ``{Flavoured jet algorithms: a comparative study},''
  \href{http://dx.doi.org/10.1007/JHEP09(2025)149}{{\em JHEP} {\bfseries 09}
  (2025) 149}, \href{http://arxiv.org/abs/2506.13449}{{\ttfamily
  arXiv:2506.13449 [hep-ph]}}.

\bibitem{NNPDF:2017mvq}
{\bfseries NNPDF} Collaboration, R.~D. Ball {\em et~al.}, ``{Parton
  distributions from high-precision collider data},''
  \href{http://dx.doi.org/10.1140/epjc/s10052-017-5199-5}{{\em Eur. Phys. J. C}
  {\bfseries 77} no.~10, (2017) 663},
  \href{http://arxiv.org/abs/1706.00428}{{\ttfamily arXiv:1706.00428
  [hep-ph]}}.

\bibitem{Hou:2019efy}
T.-J. Hou {\em et~al.}, ``{New CTEQ global analysis of quantum chromodynamics
  with high-precision data from the LHC},''
  \href{http://dx.doi.org/10.1103/PhysRevD.103.014013}{{\em Phys. Rev. D}
  {\bfseries 103} no.~1, (2021) 014013},
  \href{http://arxiv.org/abs/1912.10053}{{\ttfamily arXiv:1912.10053
  [hep-ph]}}.

\end{thebibliography}

\providecommand{\href}[2]{#2}\begingroup\raggedright\endgroup

\end{document}